\documentclass{elsarticle}
\usepackage[utf8]{inputenc}
\usepackage[T1]{fontenc}
\usepackage{fixltx2e}
\usepackage{graphicx}
\usepackage{longtable}
\usepackage{float}
\usepackage{wrapfig}
\usepackage{rotating}
\usepackage[normalem]{ulem}
\usepackage{textcomp}
\usepackage{marvosym}
\usepackage{wasysym}
\usepackage{amssymb}
\usepackage{hyperref}
\usepackage{tabularx} 
\usepackage{booktabs} 
\usepackage[section]{placeins}
\newcommand{\rk}[1]{\left(#1\right)}

\usepackage{amssymb}
\usepackage{subfig} 
\usepackage{tikz} 
\usepackage{pgfplots} 
\tikzstyle{solid}=                   [dash pattern=]
\tikzstyle{dotted}=                  [dash pattern=on \pgflinewidth off 2pt]
\tikzstyle{densely dotted}=          [dash pattern=on \pgflinewidth off 1pt]
\tikzstyle{loosely dotted}=          [dash pattern=on \pgflinewidth off 4pt]
\tikzstyle{dashed}=                  [dash pattern=on 3pt off 3pt]
\tikzstyle{densely dashed}=          [dash pattern=on 3pt off 2pt]
\tikzstyle{loosely dashed}=          [dash pattern=on 3pt off 6pt]
\tikzstyle{dashdotted}=              [dash pattern=on 3pt off 2pt on \the\pgflinewidth off 2pt]
\tikzstyle{densely dashdotted}=      [dash pattern=on 3pt off 1pt on \the\pgflinewidth off 1pt]
\tikzstyle{loosely dashdotted}=      [dash pattern=on 3pt off 4pt on \the\pgflinewidth off 4pt]
\tikzset{mark size=1}

\title{On the origin of impinging tones at low supersonic flow}

\author[cfd]{R.~Wilke\corref{cor1}}
\ead{robert.wilke@tnt.tu-berlin.de}
\author[cfd]{J.~Sesterhenn\corref{cor2}\fnref{fn1}}
\ead{joern.sesterhenn@tu-berlin.de}
\ead[url]{http://www.cfd.tu-berlin.de}

\cortext[cor1]{Corresponding author}
\cortext[cor2]{Principal corresponding author}

\fntext[fn1]{joern.sesterhenn@tu-berlin.de}

\address[cfd]{TU Berlin, Institute of Fluid Dynamics and Technical Acoustics, Müller-Breslau-Str. 12, 10623 Berlin, Germany}

\begin{document}

\begin{abstract}

Impinging compressible jets may cause deafness and material fatigue due to immensely loud tonal noise. It is generally accepted that a feedback mechanism similar to the screech feedback loop is responsible for impinging tones. The close of the loop remained unclear. One hypothesis hold up in the literature explains the emanated sound with the direct interaction of vortices and the wall. Other explanations name the standoff shock oscillations as the origin of the tones. Using direct numerical simulations (DNS) we were able to identify the source mechanism for under-expanded impinging jets with a nozzle pressure ratio (NPR) of 2.15 and a plate distance of 5 diameters. We found two different types of interactions between vortices and shocks to be responsible for the generation of the impinging tones. They are not related to screech.
\end{abstract}


\maketitle

\section{Introduction}
\label{sec-1}
A jet impinging on a flat plate may emanate incredibly loud tonal noise if the Mach number is sufficiently high $(M \gtrsim 0.7)$ and the plate is less than about 7.5 diameters away from the nozzle \cite{HoNosseir1981}. In addition to the discrete tones, the presence of the impinging plate increases the overall sound pressure level (OASPL). Marsh \cite{Marsh1961} observed that for subsonic impinging jets the OASPL increases with decreasing nozzle-to-plate distance ($h/D$).

The loud tonal components in the sound spectrum (impinging tones) were early found to be due to a feedback loop involving a shear layer instability travelling downstream and some acoustic wave travelling upstream in some, necessarily subsonic part of the flow \cite{RockwellNaudascher1979}. The same idea was convincingly applied by Ho and Nosseir 1981 \cite{HoNosseir1981} as well as Henderson and Powell \cite{HendersonPowell1993,Henderson2002}, but it remained unclear who are the culprits for the feedback loop at the wall. Ho and Nosseir identified primary vortices impinging on the wall as a possible link in the feedback chain. Powell and Henderson on the contrary identified standoff shock oscillations as the responsible mechanism within the loop.

Henderson and Powell \cite{HendersonPowell1993,Henderson2002} reported a \textit{zone of silence}: depending on the nozzle pressure ratio (NPR) and the nozzle-to-plate distance, some configurations do not allow the production of impinging tones. The analysed NPR ranges from 3.38 to 4.50. In contrast, for ideally expanded jets Krothapalli \cite{KrothapalliRajkuperan1999} et al. found continuously tones for nozzle-to-plate distances up to 10 diameters. Henderson \cite{Henderson2002} argued that both configurations (ideally and under-expanded) differ strongly in the shock-wave structure and therefore the zone of silence and the production of impinging tones must be affected by the shock-wave structure. He also proposed, that tones generated at $5 \leq h/D \leq10$ may be related to jet screech. Sinibali et al. \cite{SinibaldiMarino2015} conducted acoustic and PIV measurements of supersonic impinging jets. Nozzle pressure ratios between two and four were analysed for nozzle-to-plate distances of two, three and four diameters. The zone of silence shifts to higher values of NPR with increasing nozzle-to-plate distances ($h/D$). For $h/D=4$ the zone of silence ranges from $3.25 \leq$ NPR $\leq 4$. Sinibaldi et al. suggest that the interaction of the shear layer vortices with the wall is the only source of impinging tones in the pre-silence region, since the standoff shock is not present. In the post-silent region, the standoff shock oscillations are named as the only possible sources of the impinging tones. This is antithetical to the observations of Mitchell et al. \cite{MitchellHonnereySoria2012} and Buchmann et al. \cite{BuchmannMitchellSoria2011}, who were able to capture images of the receptivity at the nozzle by means of schlieren images from a high-speed camera. The investigated case lays in the pre-silence region (NPR$=3.2, h/D=4$) and clearly shows the presence of a standoff shock. Also Hirata et al. \cite{HirataKukita1971} observed standoff shock oscillations for large $h/D$.

Summing up, the generation of impinging tones is generally accepted to be due to a feedback mechanism. If the vortices impinging on the plate or the standoff shock oscillations generate the feedback wave is controversial and not presently clarified. Using direct numerical simulations, we are able to identify the sound source mechanism of the impinging jet in the pre-silence region for at least NPR$=2.15$ and $h/D=5$. We expect this result to hold for low NPR and sufficiently high $h/D$.

Our line of argumentation is as follows: First we shortly review some important characteristics of the free jet (section \ref{sec:jet-modes}), since there is a similar mechanism who produces tones, referred to as screech via a feedback loop mechanism. This phenomenon involves the modes of the jet. Then we shortly review what is known about the modes of the impinging jet (section \ref{sec:impinging-jet}). In section \ref{sec:flow} we describe the flow of the impinging jet using our DNS data. This includes a modal analysis focused on the frequency that appears as impinging tone. In addition, the behaviour of the two main actors (standoff shock and jet instability) are described separately. The influence of the Reynolds number and the ambient temperature are discussed. 

Section \ref{sec:sound_source_mechan} contains the main argument and the crucial message of the article: Two different sound source mechanisms exist. Sound waves are emitted either by shock-vortex- or shock-vortex-shock-interactions.

The shock-vortex-interaction is similar to screech in free shear layers but differs significantly as the shock involved is the standoff shock ahead of the wall and not part of the shock cell structure. 

Shock-vortex-shock-interaction is entirely new and can in short be described as the quenching of the sonic line in between two standoff shocks by the passing vortex. 

Both mechanism are brought into accordance with the mode of the impinging jet and the feedback loop by direct observation as well as identification of dynamic modes. Ultimately we discuss the sound spectra, why this is not screech and the zone of silence.


\section{Free jet modes and screech}
\label{sec:jet-modes}

\begin{figure}
  \centering
  \includegraphics[width=0.5\textwidth]{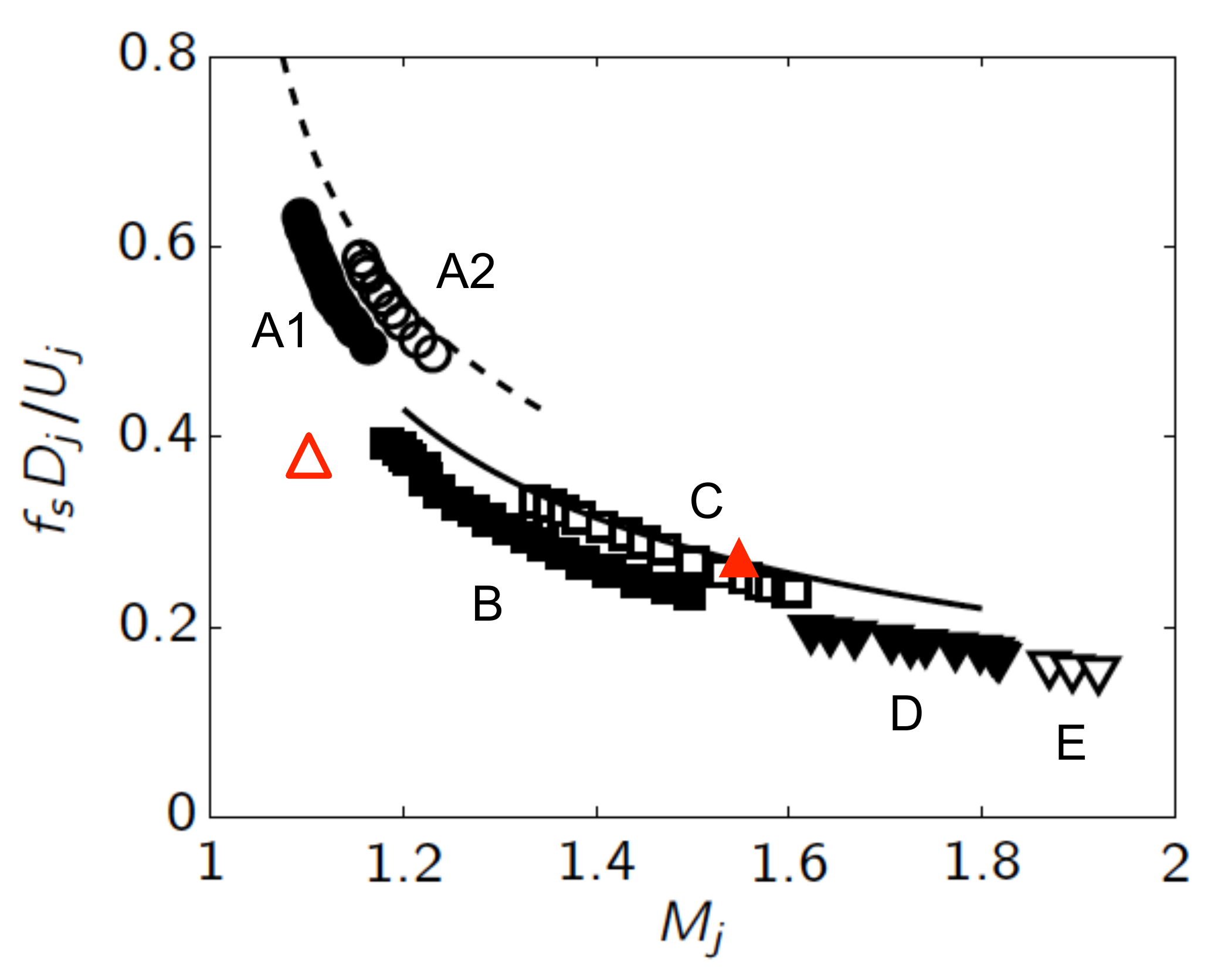}
  \caption{Supersonic jet modes: dominant screech frequency as a function of $M_j$. Adapted from \cite{Schulze2013}. The experimental data is from Panda et al. 1997 \cite{PandaRaman1997}. The two triangles with the peak oriented to the top indicate direct numerical simulations of four own group. Filled: $Re=5000, M_j=1.55$ \cite{Schulze2013}. Not filled: $Re=3300, M_j=1.11$ \cite{WilkeSesterhenn2015}. $D_j, U_j, M_j$ denote the fully expanded values of the diameter, the jet velocity and the jet Mach number. The screech frequency is labelled $f_s$.}
  \label{fig:SupersonicJetModes}
\end{figure}

Jets feature a wealth of different modes, even in the case of a round nozzle. What is important in our context is the fact that the shape of the mode determines the emanated sound. Presently not all of them are known. An overview of free chocked circular jets is given by Powell et al. \cite{PowellUmeda1992}. An impression of the different modes is given in figure (\ref{fig:SupersonicJetModes}). Depending on the fully expanded jet Mach number

\begin{equation}
	M_j=\sqrt{\frac{2}{\kappa-1} \rk{\frac{p_0}{p_{\infty}}}^{\frac{\kappa-1}{\kappa}}} \quad,
\end{equation}

we find several dominant frequencies due to screech. $M_j$ is a reasonable choice, since chocked jet have $M=1$ at the orifice, thus this Mach number is a meaningless choice. $D_j$ is the fully expanded jet diameter. For the simulations carried out at NPR$=2.15$, $D_j \approx D$ holds with a negligible error of $0.4\%$. More significant is the choice of the physical $D$ or displaced diameter $D^*$ (eq. \ref{eq:D_st}), which is discussed in section \ref{sec:sound}. But what interests us presently in the graph is the fact that we find modes $A_1$, $A_2$, $B$, $C$, $D$ and $E$. $A_1$ and $A_2$ are axisymmetric. B and D are flapping. Powell at al. \cite{PowellUmeda1992} denote them ``primarily flapping'' as they can occasionally appear as helical. A flapping mode can be considered as a superposition of two helical modes with the same amplitude and opposite sense. In case one of them is weak or missing, what is possible since B and D are not very robust, the resulting mode remains helical. Mode $C$ is helical. Mode $E$ is unknown, but in the case of an elliptic nozzle it is known to split up in different modes denoted $E_1$, $E_2$ and $E_3$. The fact that it splits up in the elliptical case could indicate that we deal with several modes indeed.

Figure \ref{fig:SupersonicJetModes} describes the influence of the Mach number only. Little is known about Reynolds number effects. The primary cause for that is a lack of data and the fact that most experiments have been performed at high Reynolds numbers, naturally occurring if experiments are done with reasonable size. However, DNS from our own group \cite{Schulze2013} at $M_j=1.55$ and $Re=5000$ as well as \cite{WilkeSesterhenn2015} at $Re=3300$ and $M_j=1.11$ indicate that the correspondence of $M_j$ to modal structure might be distorted. From figure \ref{fig:SupersonicJetModes}, based on the measurements of Panda et al. 1997 \cite{PandaRaman1997}, one would expect a torodial mode appearing at $M_j = 1.11$. However, our simulation with a cold jet (ambient temperature $T_{\infty} = 373.15$ K and total inlet temperature $T_0 = 293.15$ K) features a helical mode corresponding to a Strouhal number of $Sr = 0.375$. This frequency fits into the range of mode B, which can be helical as well. This is also supported by the computation of Sesterhenn et al. \cite{SesterhennMiranda2013} who report a change in modal structure when particles are added to the jet. This might be due to the change of the density of the jet, which then would also lead to the conclusion that heating changes the modes.

One more issue deserves some special emphasis: A closer look at figure \ref{fig:SupersonicJetModes} shows that for some Mach numbers, more than one possible modes exist. For example at $M \approx 1.2$, we observe $A_1$, $A_2$, and $B$ as possible candidates, each of which having a different frequency. The mechanism of the mode selection is unclear and it would be worthwhile trying to force the jet in one or another mode. Given identical boundary conditions, a mode selection in form of initial conditions must come in. We do not discuss this issue further, but underline the different coexisting states.

The specific sound source in supersonic jets that depends on the described modes is referred to as screech. Screech is a mark for discrete tones that are generated by a feedback mechanism: Vortical structures develop in the shear layer of the jet and grow while they are convected downstream. When the large scale structures reach the fourth or fifth shock cell, both interact and emit strong acoustic waves that propagate upstream. These reach the nozzle lip or upper plate and excite the shear layer of the jet which leads to new instability waves and the close of the feedback mechanism.

The interaction was described by Suzuki and Lele \cite{SuzukiLele2003}, based on a two-dimensional DNS. Fernandez and Sesterhenn \cite{FernandezSesterhenn2015} performed a three-dimensional DNS of a round starting jet. They found that the shock-wave present in the core of the trailing jet is bent by
the vortices from the shear layer that reach the shock-wave. As a result, the shock transforms into a strong acoustic wave that is radiated into the outer region. This phenomenon is very similar to the shock-vortex-interaction, as described in section \ref{sec:sv}. However, the involved shocks are different: In the free jet, the shock diamond interacts with the vortices, whereas in the impinging jet it is the moving standoff shock.

\section{Impinging jet modes}
\label{sec:impinging-jet}

As explained in the previous section, free jet screech is strongly connected with the modal structure of the jet. Despite this topic is still being investigated, the gained knowledge during the past decades is considerable. In contrast, comparably little is known about the impinging jet modes.

In \cite{TamAhuja1990} Tam and Ahuja argue that only axisymmetrical modes are possible for subsonic impinging jets, whereas also helical modes can occur in the supersonic case. This statement is based on an analytical model and the studies found in literature: Neuwirth \cite{Neuwerth1974,Neuwerth1981} observed axisymmetrical, helical and flapping (superposition of two helical) modes for supersonic impinging jets. Additionally, a helical coherent structure for a free jet was observed at $M=0.8$. Adding an impinging plate (without changing any other parameter), the mode changed to axisymmetrical. Nosseir and Ho \cite{NosseirHo1982} likewise observed an axisymmetrical mode at $M=0.7$ for an impinging jet.

Krothapallo et al. \cite{KrothapalliRajkuperan1999} conducted an experiment involving ideally expanded free and impinging jets at $M_j=1.5$. The mode of the free jet is helical. Approaching the plate, this mode stays dominant until $h/D=8$. Between $h/D=4$ and 6, the axisymmetrical begins to dominate. A further decrease of $h/D$ leads to a re-emergence of the helical mode.

Tsubokura et al. \cite{TsubokuraKobayashi2003} conducted a DNS with a Reynolds number of 2000. This flow is not fully turbulent, since therefore Reynolds numbers above about 3000 are required. The simulation ($h/D=10$) showed a mode that is axisymmetrical close to the orifice plate, but develops an asymmetry close to the impinging plate.

A recent numerical investigation was performed by Uzun et al. \cite{UzunKumar2013}. He conducted a large eddy simulation with a plate distance of five diameters and a Mach number of 1.5. The coherent axisymmetrical structures found using a DMD, correspond to the dominant tone at $Sr\approx 0.33$.\\

\section{Computational setup}
\label{sec:geometry}

The governing Navier-Stokes equations are formulated in a characteristic pressure-velocity-entropy-formulation, as described by Sesterhenn \cite{Sesterhenn2001} and are solved directly numerically. This formulation has advantages in the fields of boundary conditions, parallelization and space discretisation. No turbulence modelling is required since the smallest scales of turbulent motion are resolved. The spatial discretisation uses 6th order compact central schemes for the diffusive terms and compact 5th order upwind finite differences for the convective terms. To advance in time a 4th order Runge-Kutta scheme is applied. In order to avoid Gibbs oscillations in the vicinity of the standoff shock an adaptive shock-capturing filter developed by Bogey et al. \cite{BogeyCacqueray2009} that automatically detects shocks is used.\\

The computational domain is delimited by four non-reflecting boundary conditions, one isothermal wall which is the impinging plate and one boundary consisting of an isothermal wall and the inlet. The walls are fully acoustically reflective. The location of the nozzle is defined using a hyperbolic tangent profile with a disturbed thin laminar annular shear layer as described in \cite{WilkeSesterhenn2014}.\\
A sponge region is applied for the outlet area $r/D >5$, that smoothly forces the values of pressure, velocity and entropy to reference values. This destroys vortices before leaving the computational domain. The reference values at the outlet were obtained by a preliminary large eddy simulation of a greater domain.\\

The grid is refined in the wall-adjacent regions in order to ascertain a maximum value of the dimensionless wall distance $y^+$ of the closest grid point to the wall not larger than one for both plates. For the wall-parallel-directions a slight symmetrical grid stretching is applied, which refines the shear layer of the jet. The refinements use hyperbolic tangent respectively hyperbolic sin functions resulting in a change of the mesh spacing lower than $1\%$ for all directions and cases. The table \ref{tab_para} shows the physical and geometrical parameters of the simulations.\\

\begin{table}
	\caption{Geometrical and physical parameters of the simulation. $p_o, p_{\infty}, T_o, T_{\infty},T_W, Re, Pr, \kappa, R$ denote total- and ambient pressure, total-, ambient and wall temperature, Reynolds number, Prandtl number, ratio of specific heats and the specific gas constant.}
	\begin{tabularx}{\columnwidth}{p{2mm} XXXXXXX p{17mm}}
	\toprule
	N$^{\circ}$ & $p_o/p$ & $p_{\infty}$ & $T_o$ & $T_{\infty}=T_W$ & $Re$ & $Pr$ & $\kappa$ & $R$\\
	 & & [Pa] & [K] & [K] & && &[J/(kg K)]\\
	\midrule
    \#1 & $2.15$ & $10^5$ & $293.15$ & $373.15$ & $3300$ & $0.71$ & $1.4$ & $287$\\
    \#2 & $2.15$ & $10^5$ & $293.15$ & $293.15$ & $3300$ & $0.71$ & $1.4$ & $287$\\
    \#3 & $2.15$ & $10^5$ & $293.15$ & $293.15$ & $8000$ & $0.71$ & $1.4$ & $287$\\
	\end{tabularx}
	\begin{tabularx}{\columnwidth}{p{2mm} X p{28mm} p{6mm} XX}
	\toprule
	N$^{\circ}$ & domain size & grid points & max. $y^+$ & grid width x,z & grid width y\\
	& $[D]$ & & & $[D]$ & $[D]$\\
	\midrule
    \#1 & $12 \times 5 \times 12$ & $512 \times 512 \times 512$ & $0.67$ & $0.0199 .. 0.0588$ & $0.0017 .. 0.0159$\\
    \#2 & $12 \times 5 \times 12$ & $512 \times 512 \times 512$ & $0.77$ & $0.0199 .. 0.0588$ & $0.0017 .. 0.0159$\\
    \#3 & $12 \times 5 \times 12$ & $1024 \times 1024 \times 1024$ & $1.02$ & $0.0099 .. 0.0296$ & $0.0012 .. 0.0072$\\
	\bottomrule
	\end{tabularx}
	\label{tab_para}
\end{table}

\section{Description of the flow}
\label{sec:flow}


\subsection{Dynamic mode decomposition}
\label{sec:DMD}

A dynamic mode decomposition (DMD) is used to relate coherent structures of the flow field to the tonal noise of the supersonic impinging jet. The DMD, as described by Schmid and Sesterhenn \cite{SchmidSesterhenn2008,Schmid2011}, extracts dynamic information out of a sequence of snapshots for a specific time interval that are either generated experimentally or numerically. In our case, 120 two-dimensional snapshots of the pressure field are used. Therewith we map six period length of the cycle described in the following. The temporal dynamics of the flow is approximated by a linear snapshot to snapshot operator. The dominant eigenfunctions of this evolution operator form a set of dynamically relevant modal structures (dynamic modes). The mathematical background as well as the algorithm are given in \cite{Schmid2011}.\\

\begin{figure}
\captionsetup[subfigure]{labelformat=empty}
\centering
\subfloat[]{\includegraphics[width=0.32\textwidth]{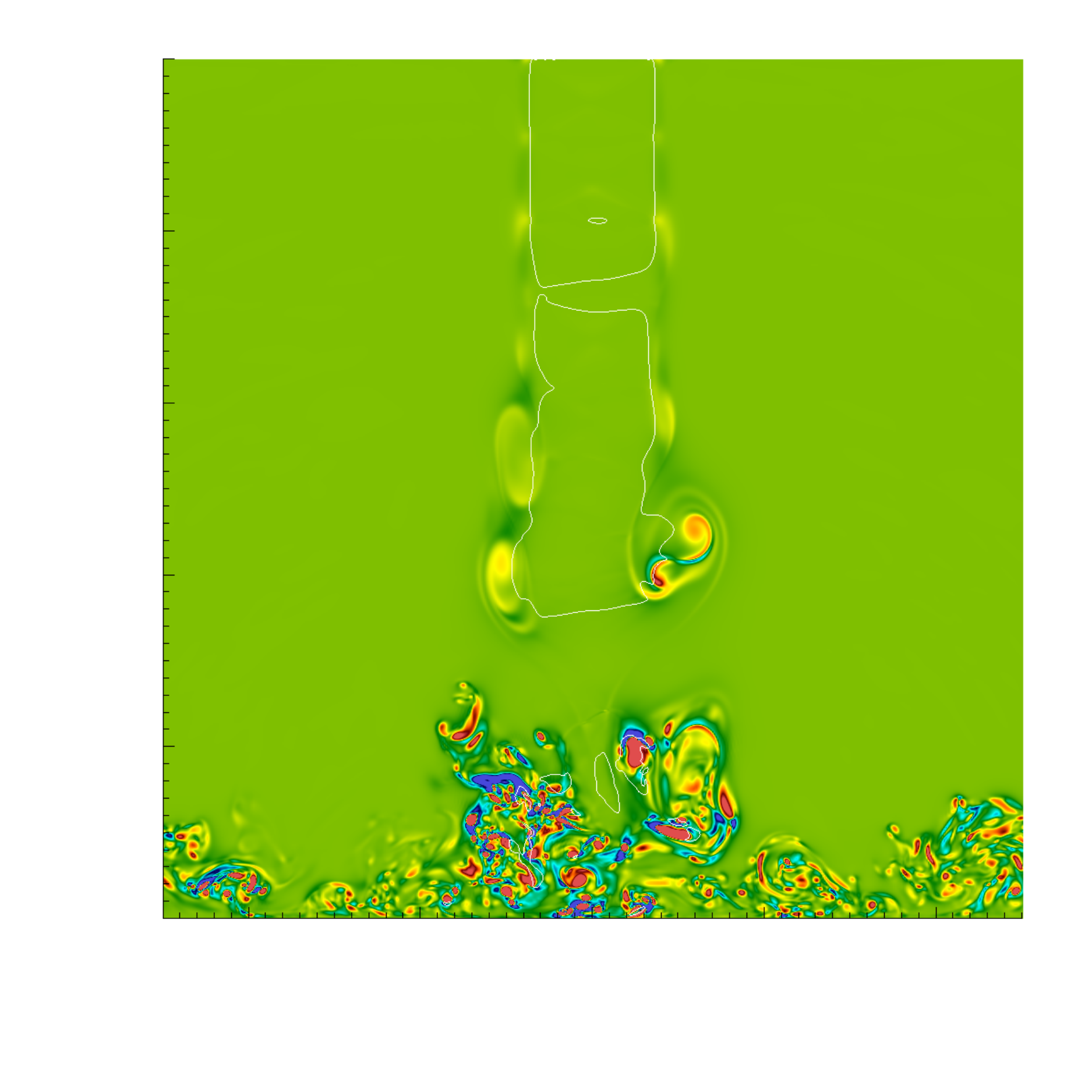}}
\subfloat[]{\includegraphics[width=0.32\textwidth]{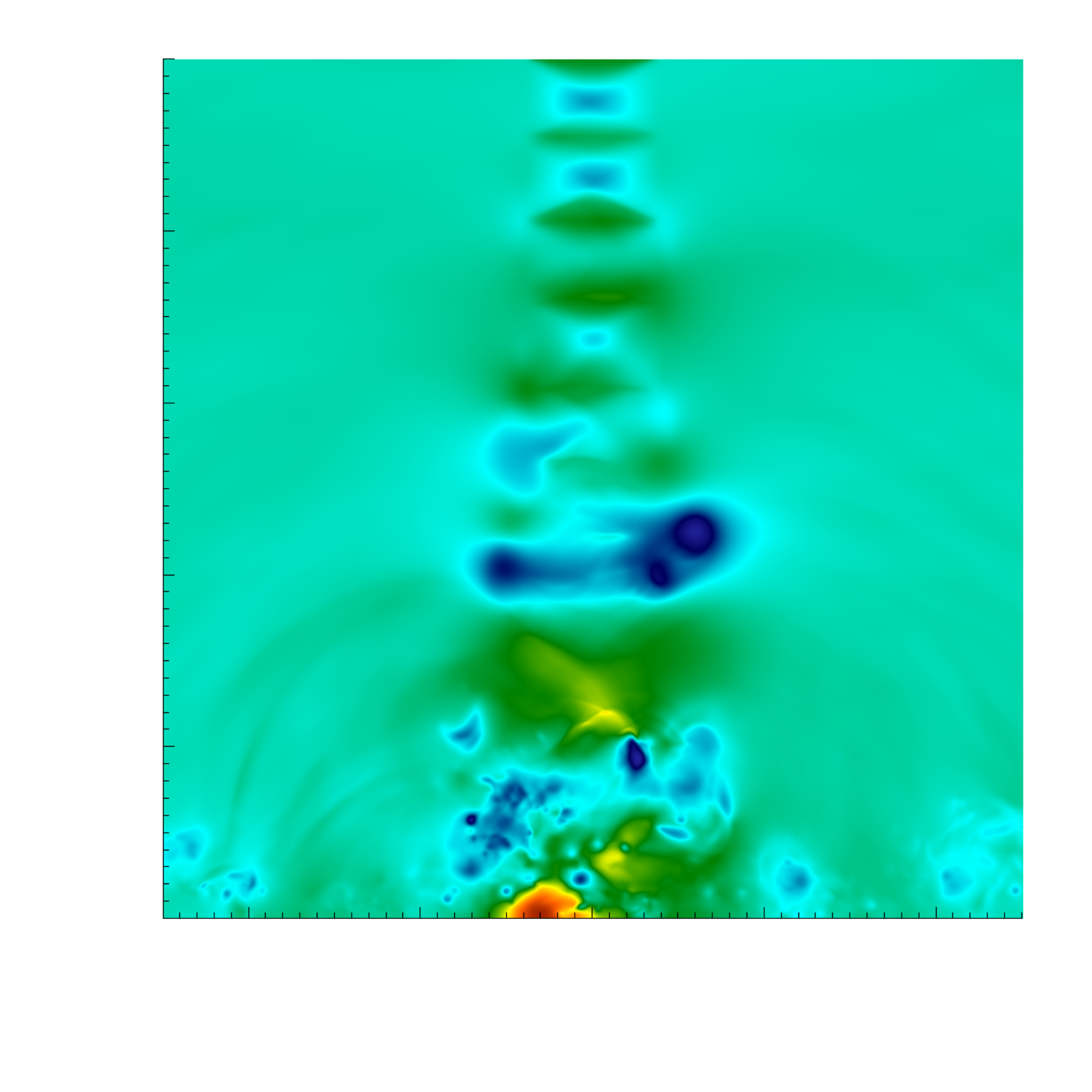}}
\subfloat[]{\includegraphics[width=0.32\textwidth]{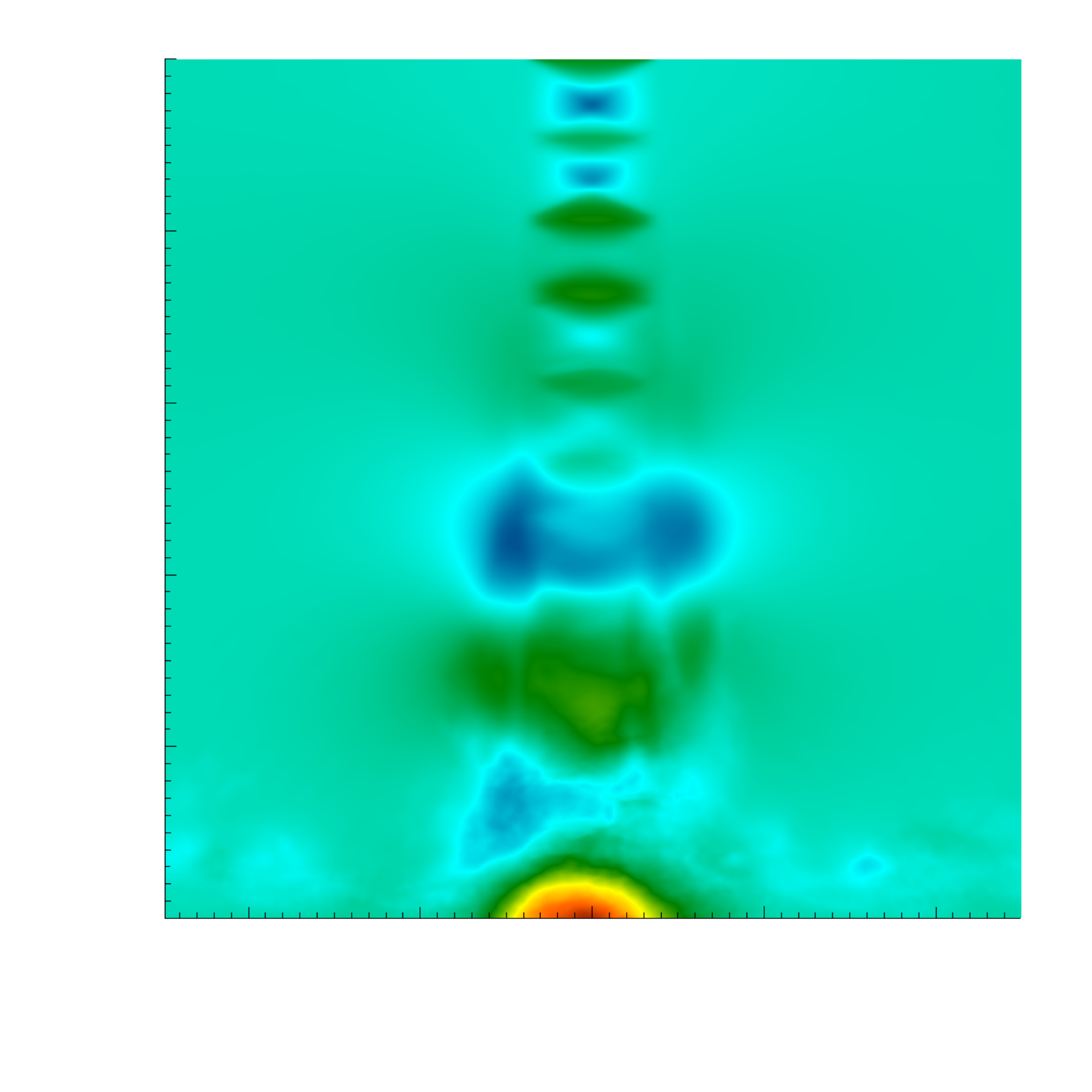}}\\[-14.5mm]

\subfloat[]{\includegraphics[width=0.32\textwidth]{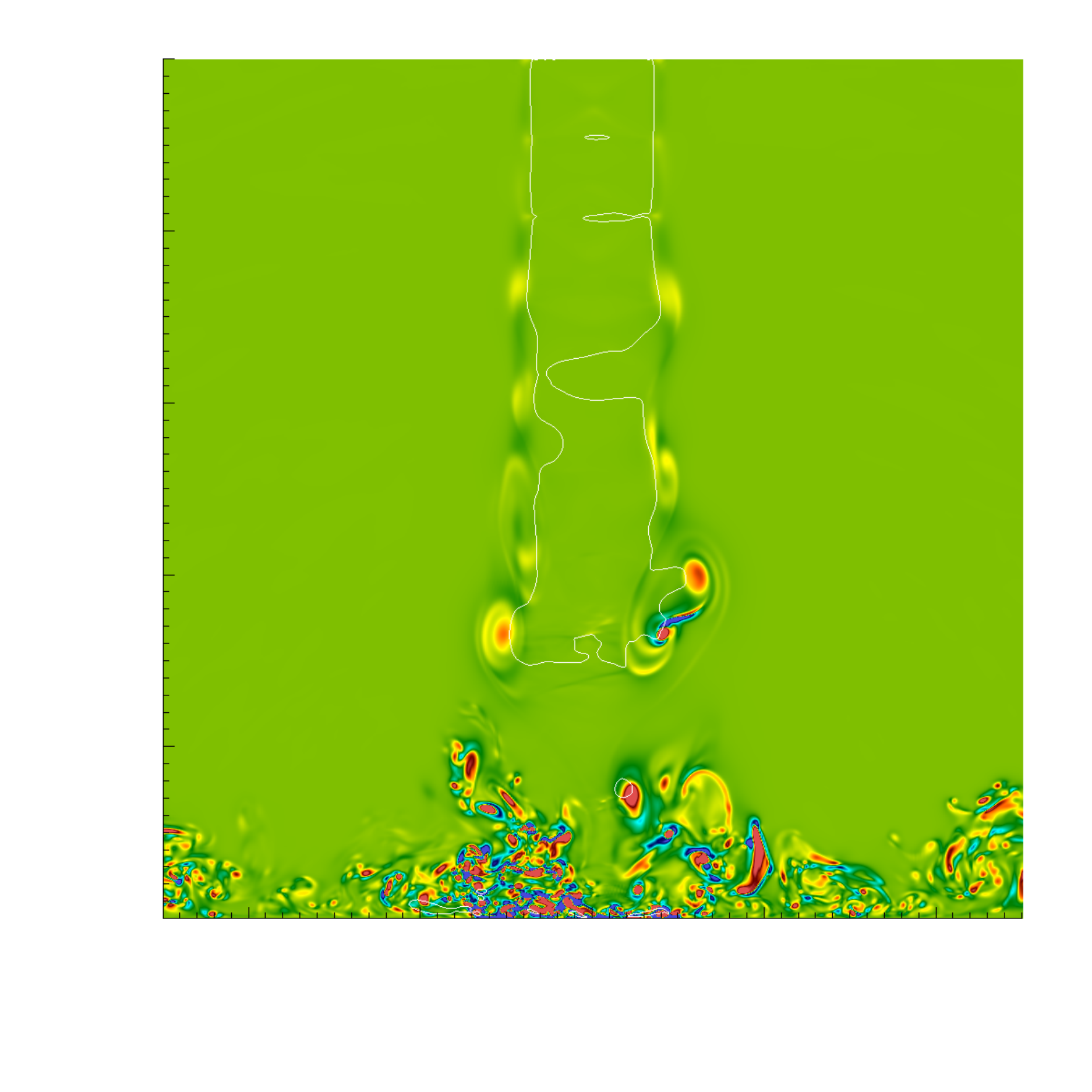}}
\subfloat[]{\includegraphics[width=0.32\textwidth]{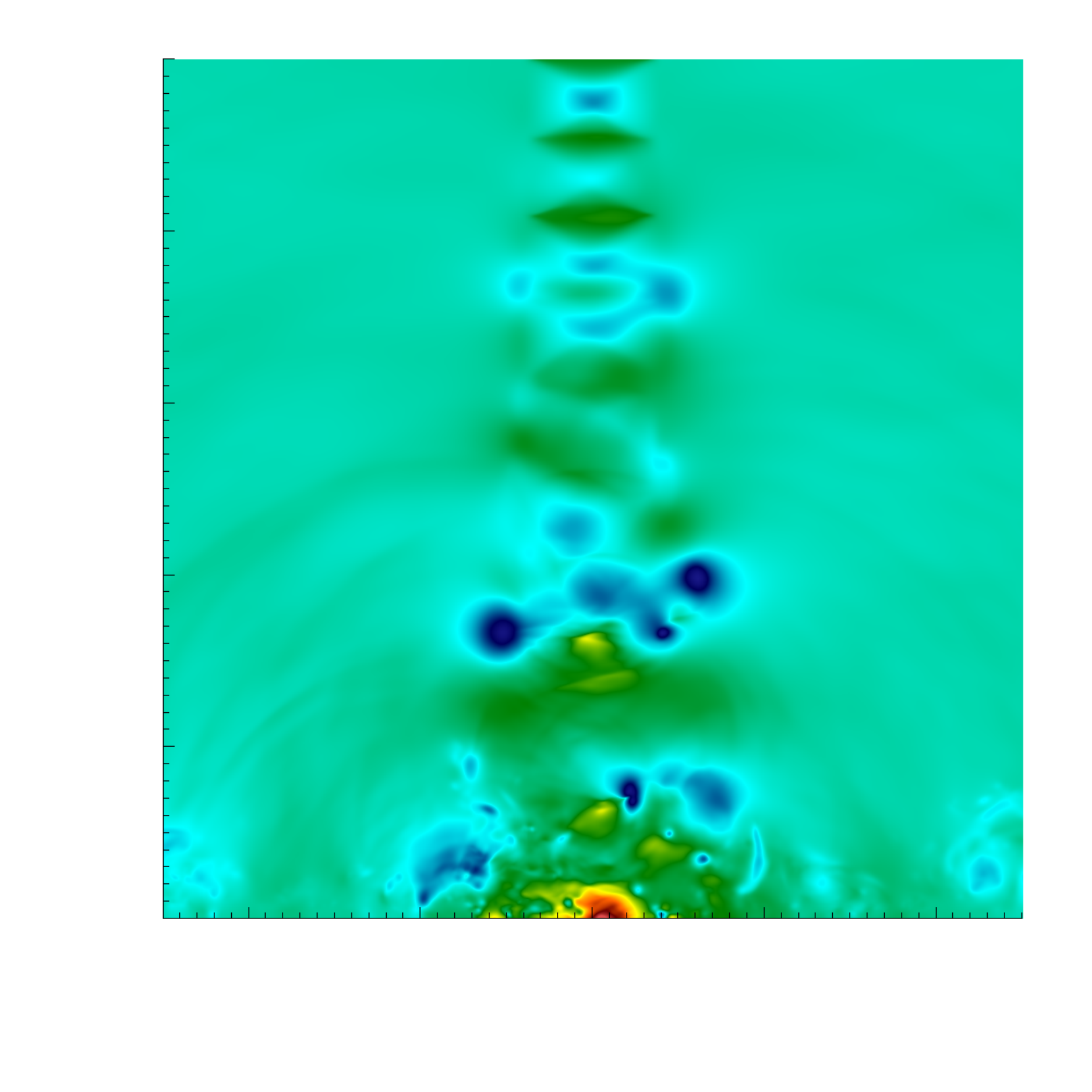}}
\subfloat[]{\includegraphics[width=0.32\textwidth]{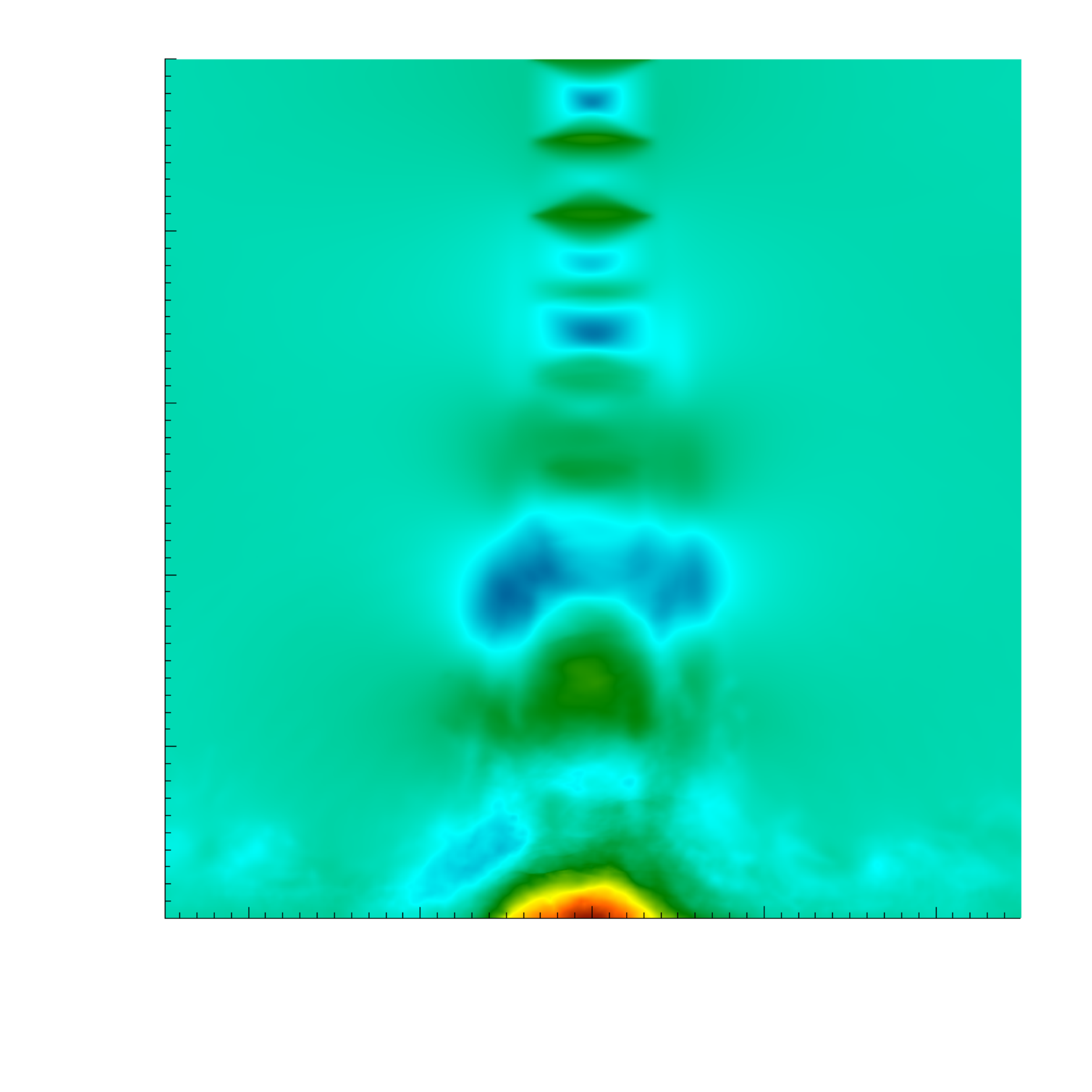}}\\[-14.5mm]

\subfloat[]{\includegraphics[width=0.32\textwidth]{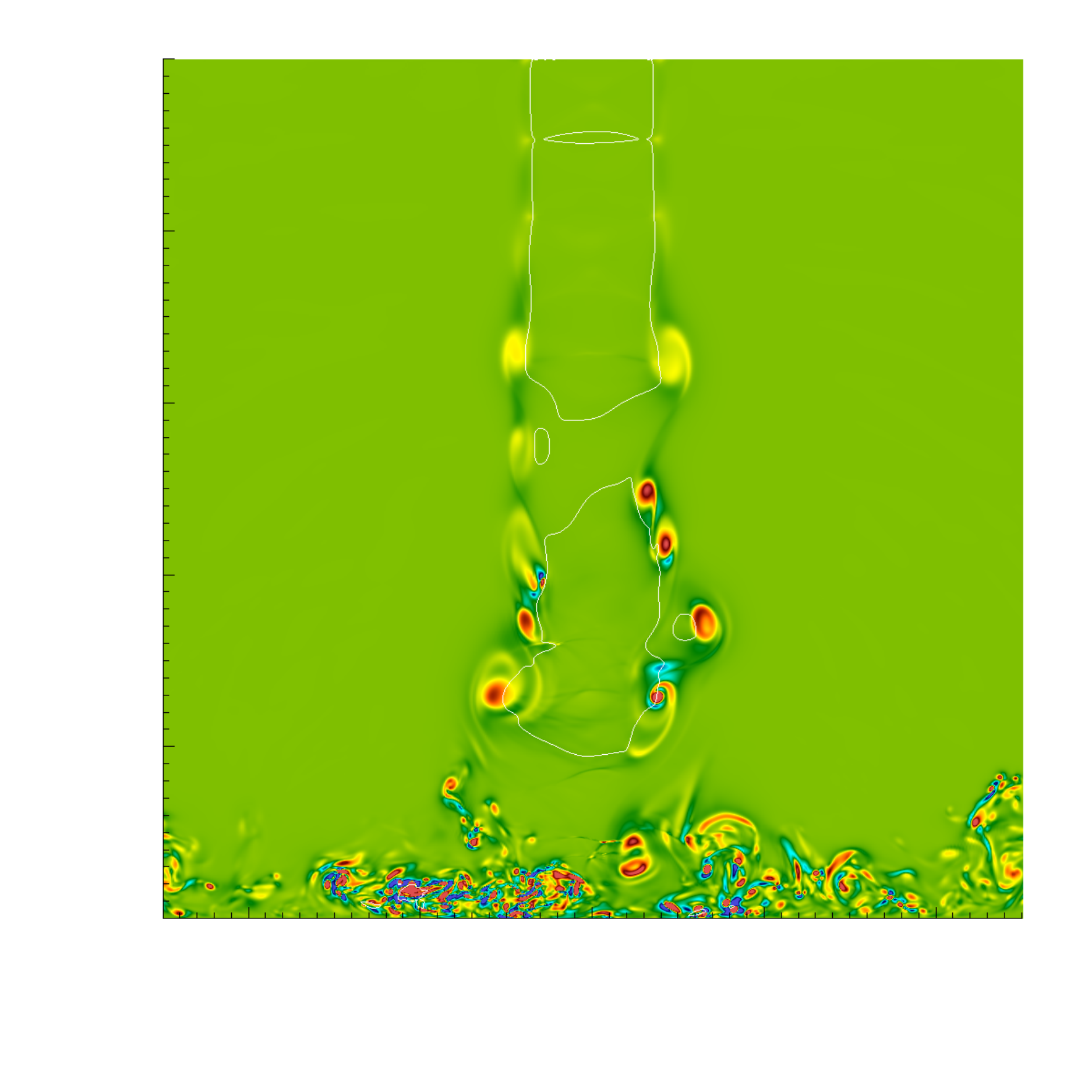}}
\subfloat[]{\includegraphics[width=0.32\textwidth]{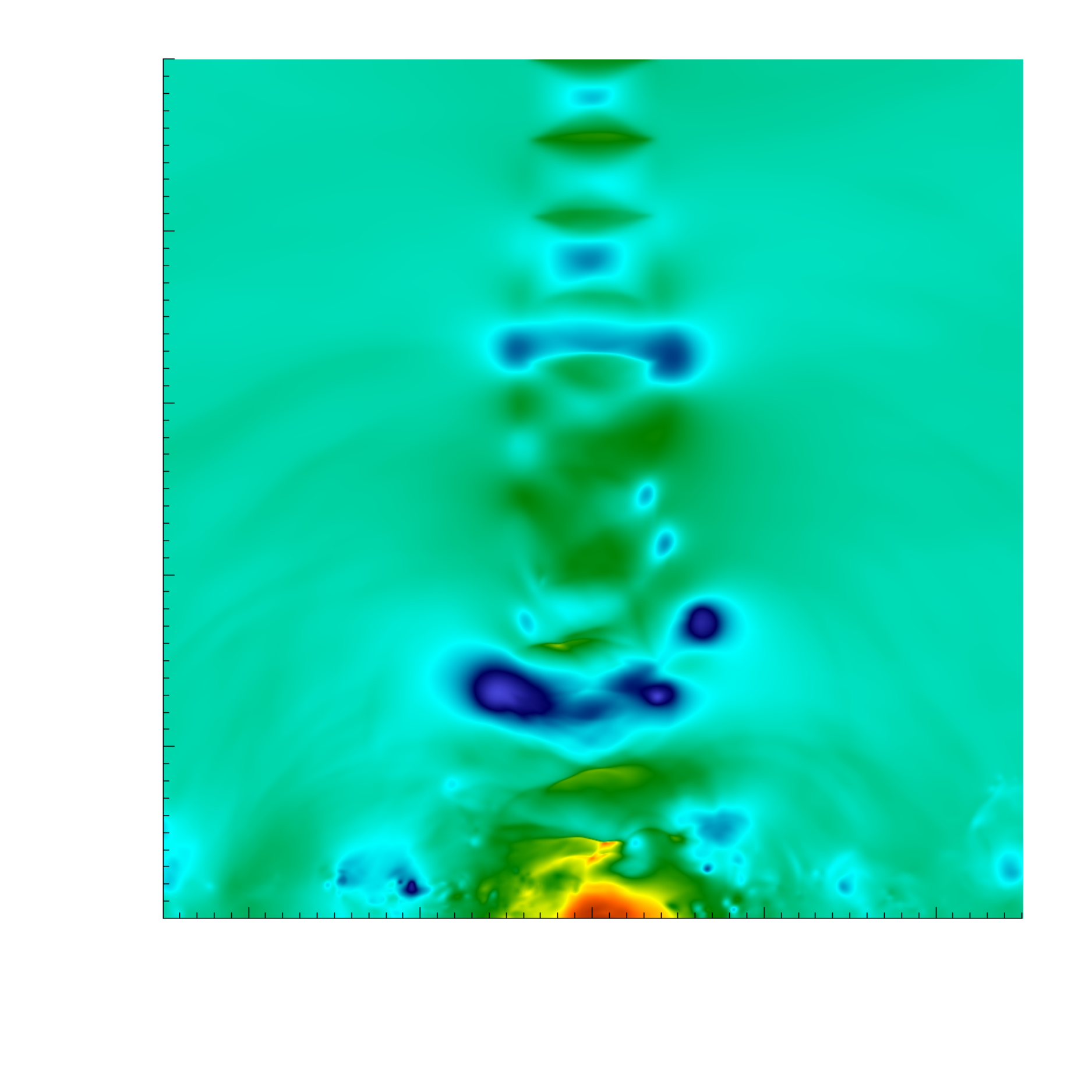}}
\subfloat[]{\includegraphics[width=0.32\textwidth]{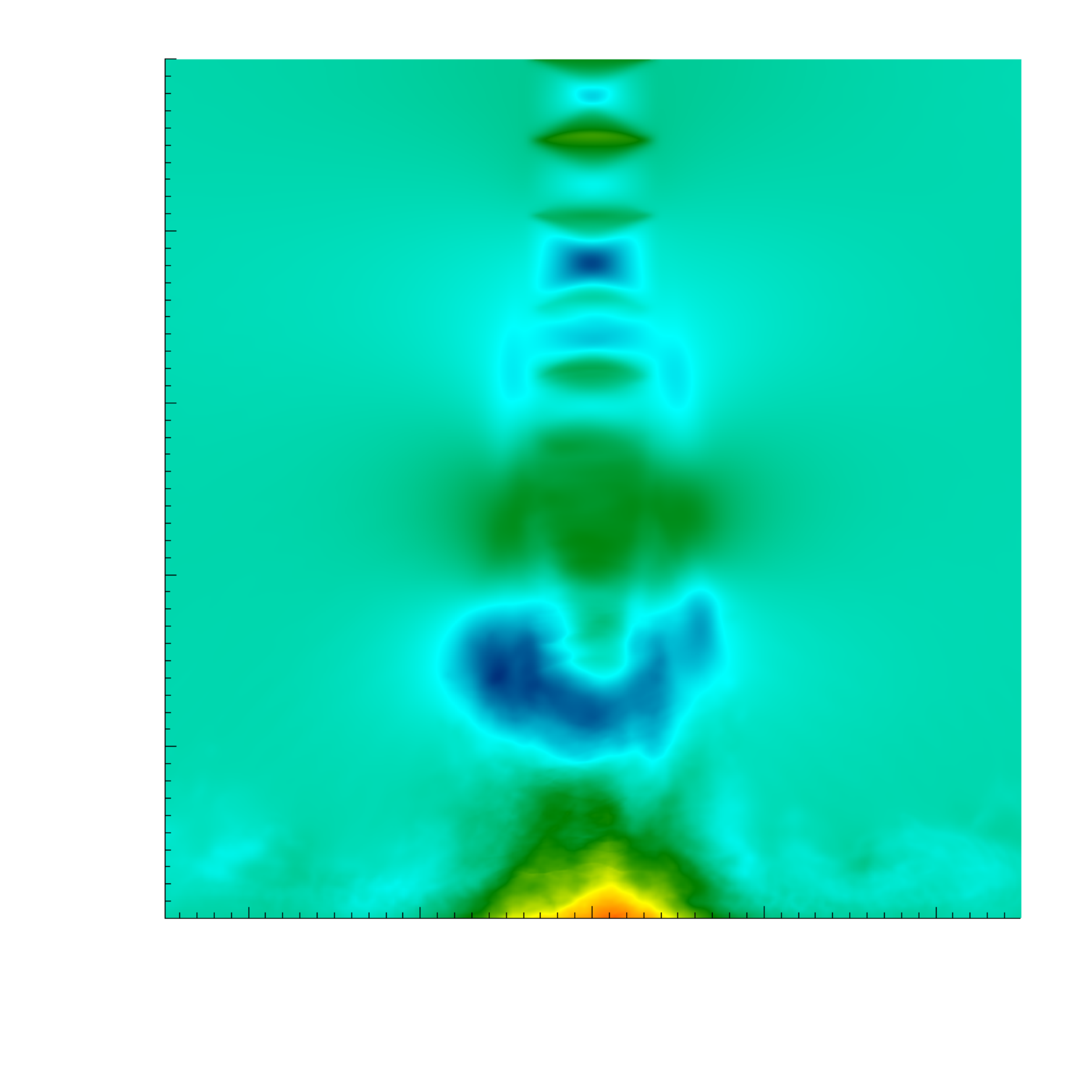}}\\[-14.5mm]

\subfloat[]{\includegraphics[width=0.32\textwidth]{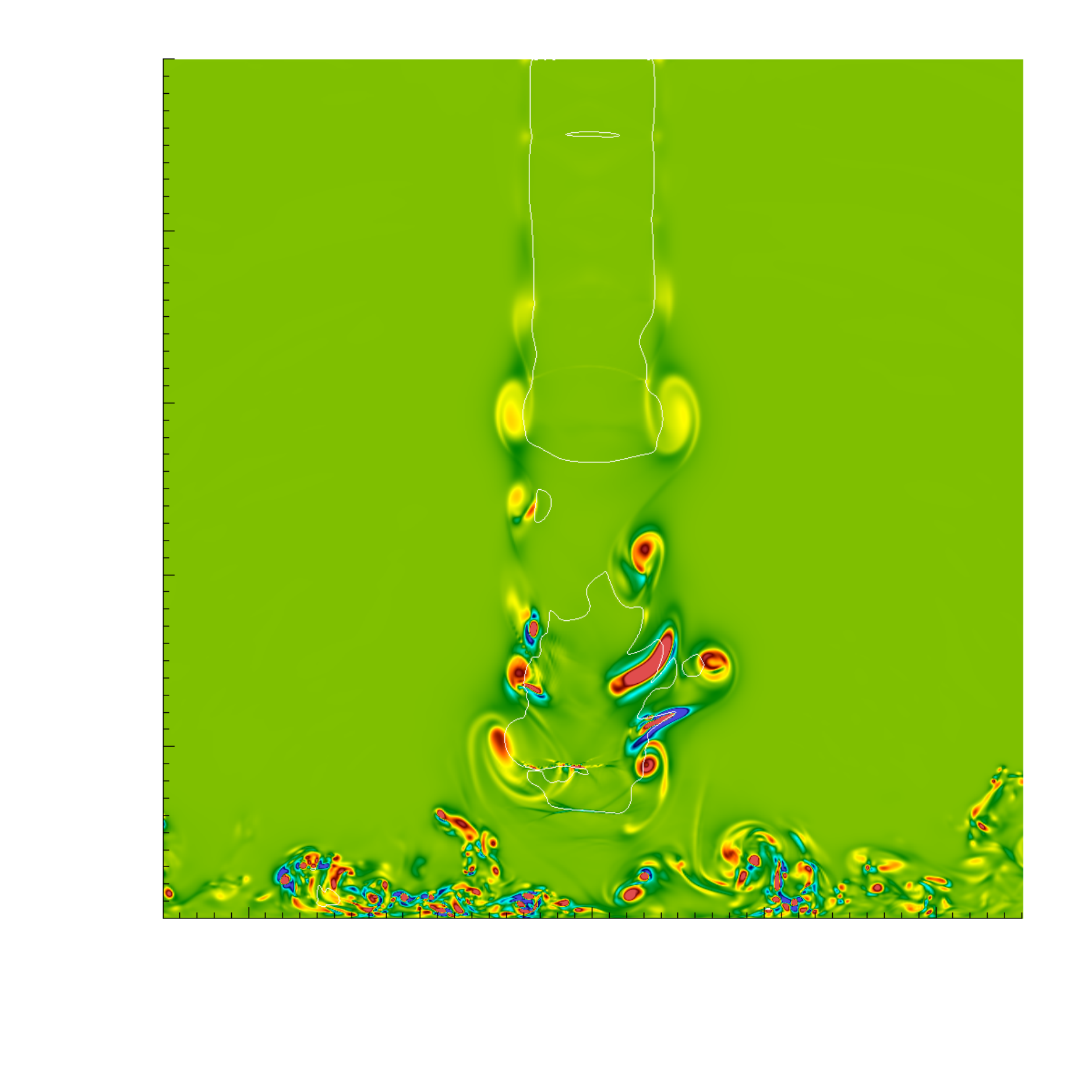}}
\subfloat[]{\includegraphics[width=0.32\textwidth]{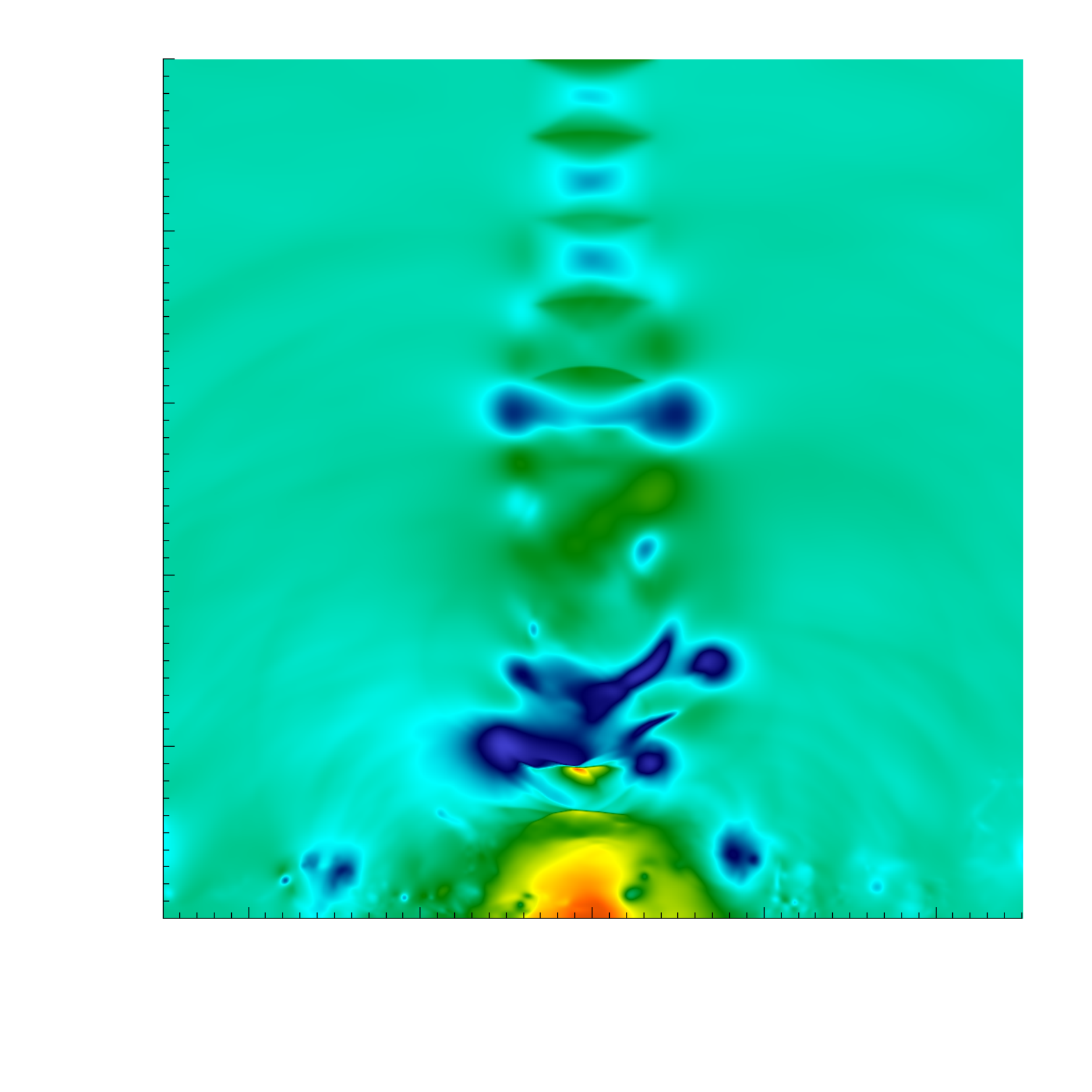}}
\subfloat[]{\includegraphics[width=0.32\textwidth]{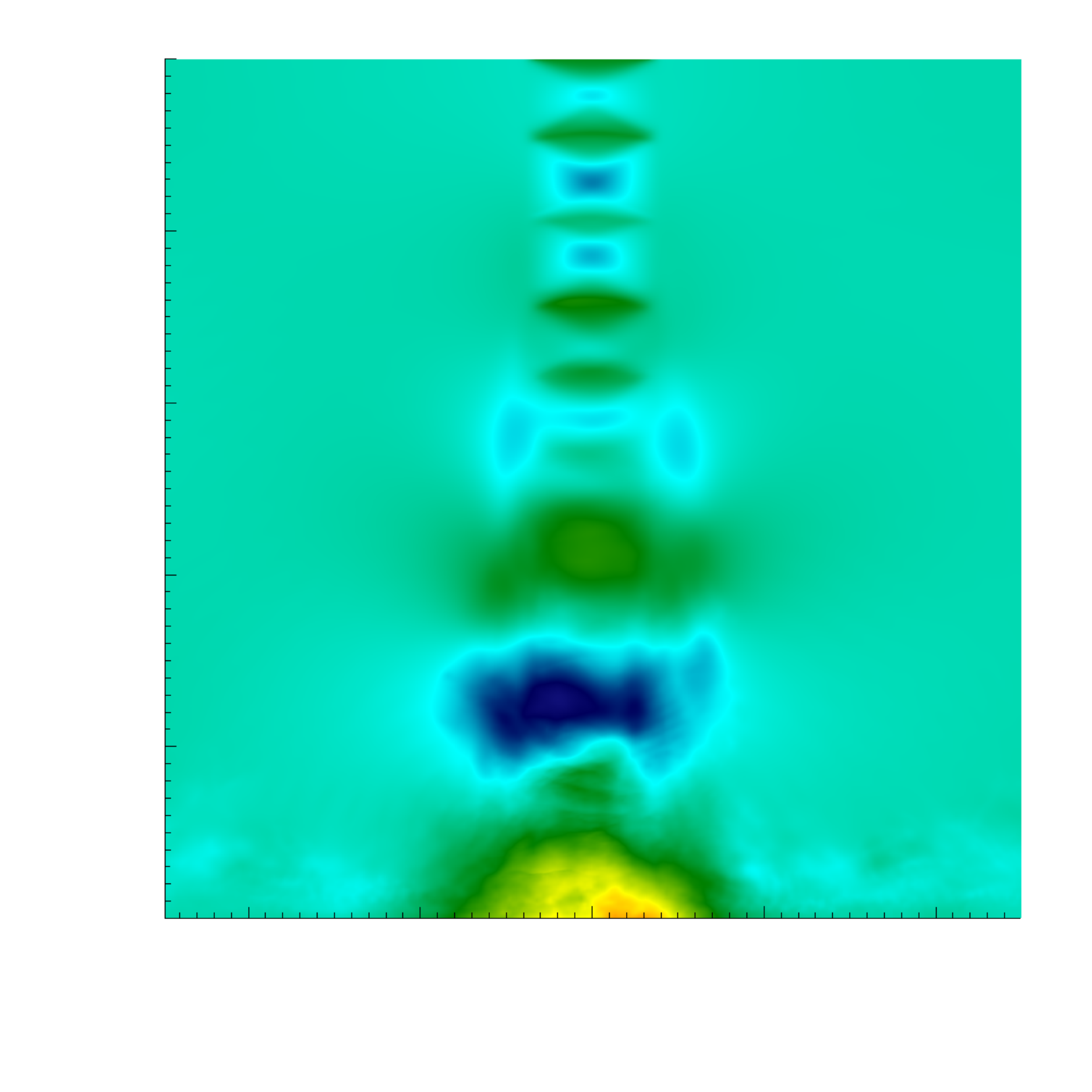}}\\[-14.5mm]

\subfloat[]{\includegraphics[width=0.32\textwidth]{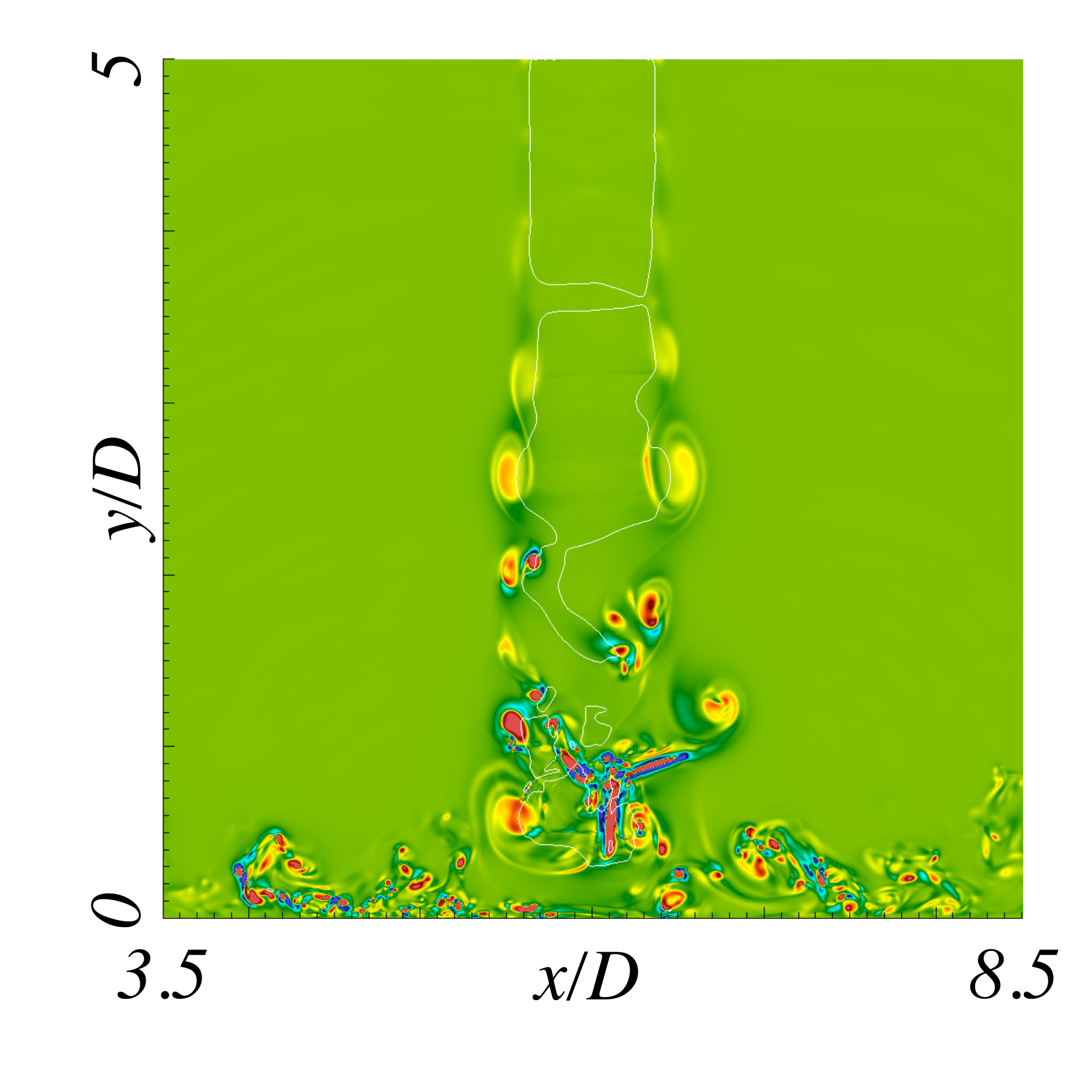}}
\subfloat[]{\includegraphics[width=0.32\textwidth]{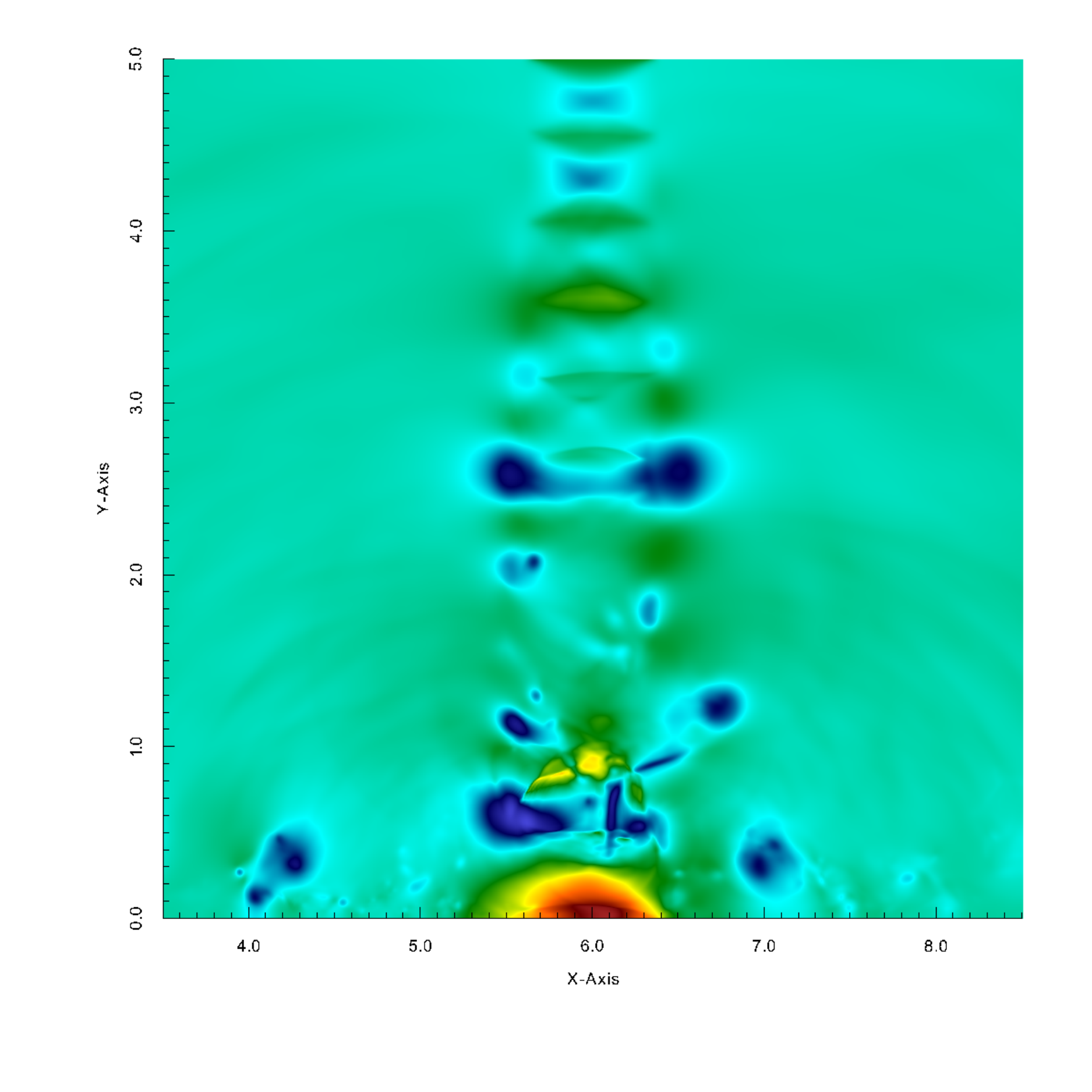}}
\subfloat[]{\includegraphics[width=0.32\textwidth]{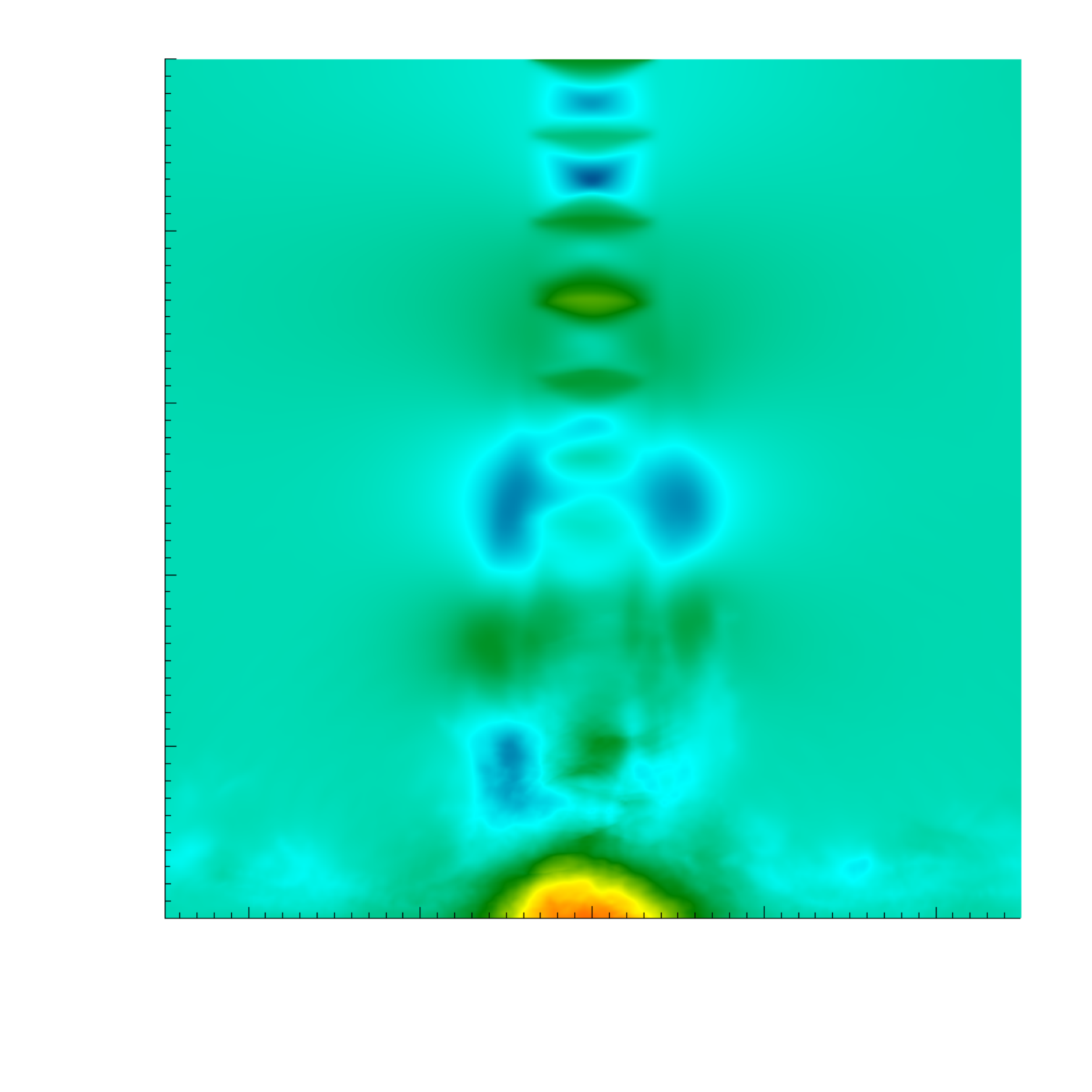}}\\[-10mm]

\subfloat[]{\includegraphics[width=0.33\textwidth]{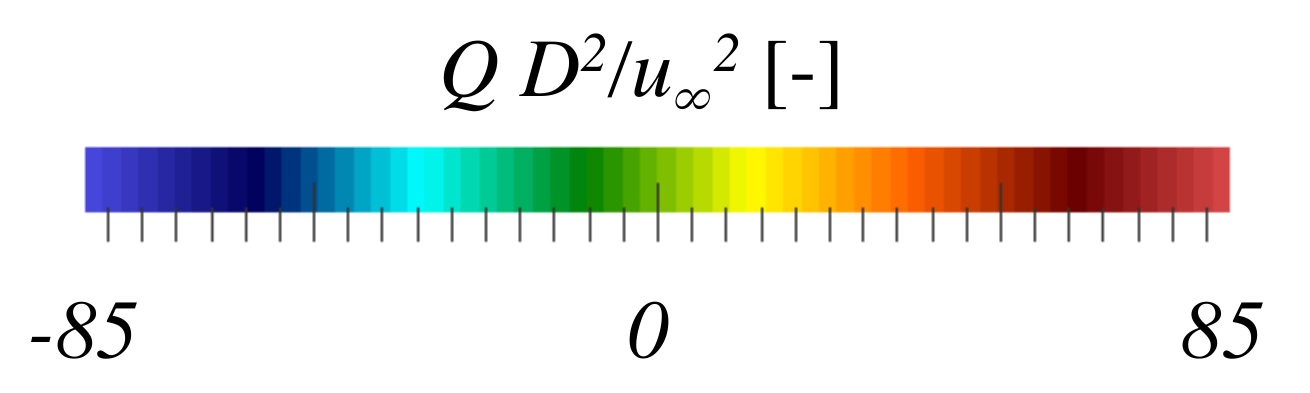}}
\hspace{0.2cm}
\subfloat[]{\includegraphics[width=0.33\textwidth]{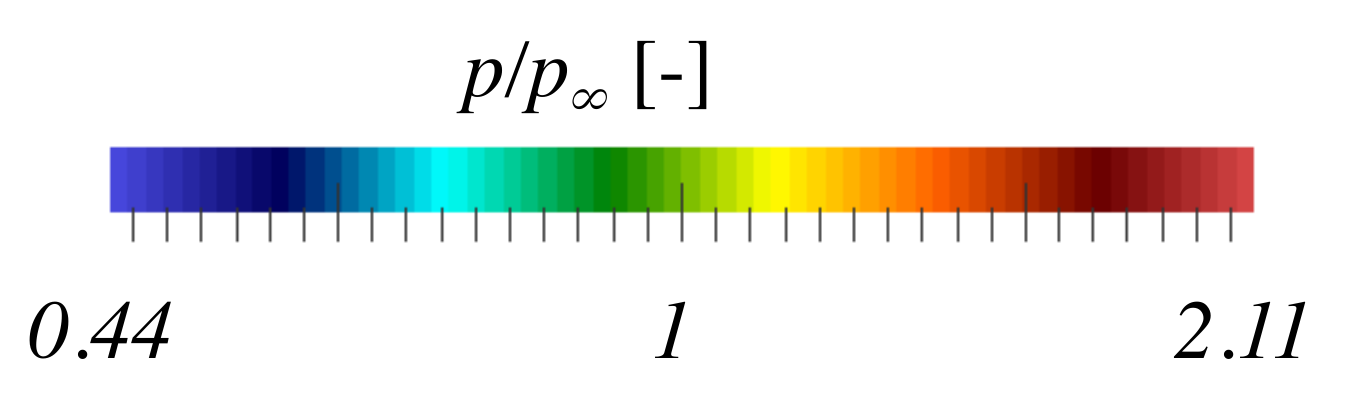}}\\[-5.5mm]

\caption{Cycle of the sound source mechanism ($Re=8000$). First column: normalised values of $Q$. Second and third column: pressure of the original flow field and of the reconstruction using the mean field and the relevant dynamic mode with a Strouhal number of $Sr=0.32$. The snapshots (rows) are in consecutive order.}
\label{fig:DMD}
\end{figure}

In order to relate the correct structures of the impinging jet to the impinging tones, we anticipate that the sound is radiated with a Strouhal number of $Sr=0.32$, which is shown later in section \ref{sec:sound}. Performing the DMD, we find a relevant mode with this frequency. In the following, a description of the flow regarding that mode is given. Simulation \#3 (see table \ref{tab_para}) with a Reynolds number of 8000 is used for this purpose. Figure \ref{fig:DMD} shows a full period of the cycle including five snapshots, one in each row. Snapshot number six, which is not shown would be again at the same phase point like the first one. In the left and middle column, the original flow field is shown ($Q$ respectively pressure $p$). The right column shows the pressure obtained from the reconstruction of the flow field using only the 0-mode, which is the time mean and the two complex conjugated dynamic modes with the frequency of the impinging tone ($Sr=0.32$). In the first point in the phase (first row) there is a highly turbulent area with plenty of small vortices close to the stagnation point ($y/D \lesssim 1$). These vortices are left from the former period and will be explained later. However the flow in this area is mainly subsonic. Large vortical structures can be found in the upper part of the domain ($y/D\gtrsim 2$). These belong to the new period that we investigate now. The first vortex ring in streamwise direction ($y/D\approx 2$), that in this specific period includes a split off (see section \ref{sec:jet_instab}), is significantly stronger than the following ones. This can becomes clear regarding the original pressure field (middle), and especially the reconstructed pressure field (right). Therefore it is referred to as head vortex. Slightly in front of the head vortex is the sonic line. Advancing in time (second and third row), the following vortices accelerate, as described in section \ref{sec:jet_instab}. This leads to a split of the supersonic area, as indicated by the sonic line. In the fourth row, the supersonic area approaches the stagnation point, encounters high pressure and forms the standoff shocks. Now shock-vortex- and the shock-vortex-shock-interactions occur, as described later in section \ref{sec:sv} and \ref{sec:svs}. As a result of the thereby produced strong pressure waves, the large structures (vortex rings) get destroyed and the supersonic area disappears. This can be seen in the last row. The breakdown of the large vortices even continues in the beginning of the new period, as shown in the first row.\\

The dynamic mode decomposition proves two statements:

\begin{itemize}
	\item The impinging tone frequency is the frequency with which a strong vortex ring (head) develops and draws in subsequent vortices leading to interactions of those structures with the standoff shock.
	\item The mode is mainly axisymmetric and not flapping or helical.
\end{itemize}

\subsection{Standoff shock}
\label{sec:standoff-shock}

As described in the previous section, standoff shocks develop due to the approach of the supersonic area to locations of high pressure ($\gtrsim 1.3 \cdot p_{\infty}$), close to the stagnation point. This happens within a periodic cycle. As a result of this cycle, the shocks are not continuously present. This can be seen in the pressure snapshots of figure \ref{fig:DMD} (middle column). In the first two rows no standoff shock can be observed. Advancing in time, the shocks develop and can be clearly seen in the fourth row at $y/D\approx 0.6$ and $0.9$. The creation of the shocks occur at the border of locations with high pressure, connected to high pressure gradients. Those are either the stagnation point or a lump which was split-off from the stagnation point due to the strong pressure waves in the previous period. However the high pressure gradients tend to move away from the stagnation point. The interaction with the contrary moving vortices is inevitable. Due to the highly turbulent flow field ($Re=8000$) the stagnation point produces multiple high pressure lumps and therefore multiple standoff shocks.


\subsection{Jet instability}
\label{sec:jet_instab}
\begin{figure}
\captionsetup[subfigure]{labelformat=empty}
\centering
\subfloat[]{\includegraphics[width=0.33\textwidth]{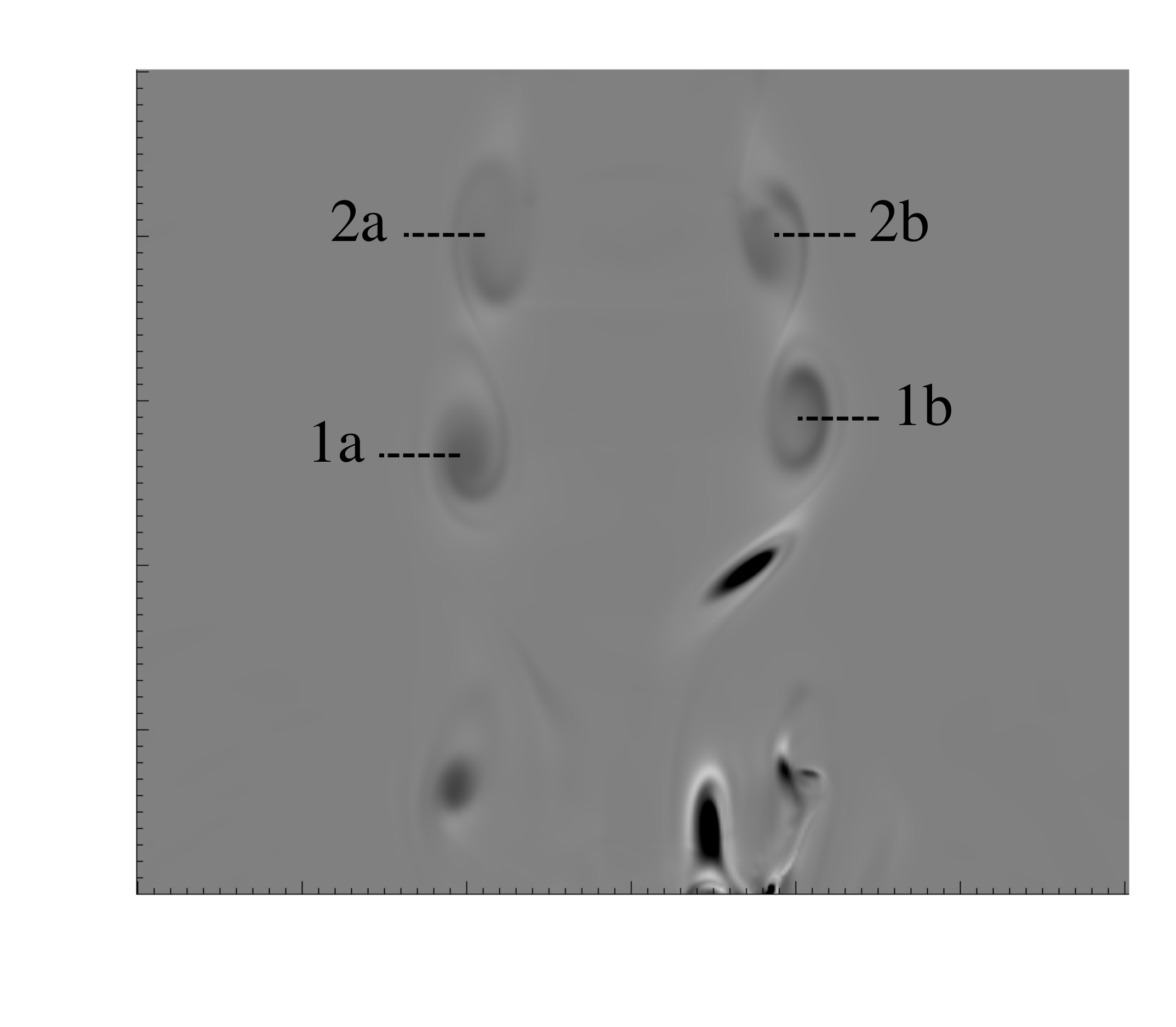}}
\subfloat[]{\includegraphics[width=0.33\textwidth]{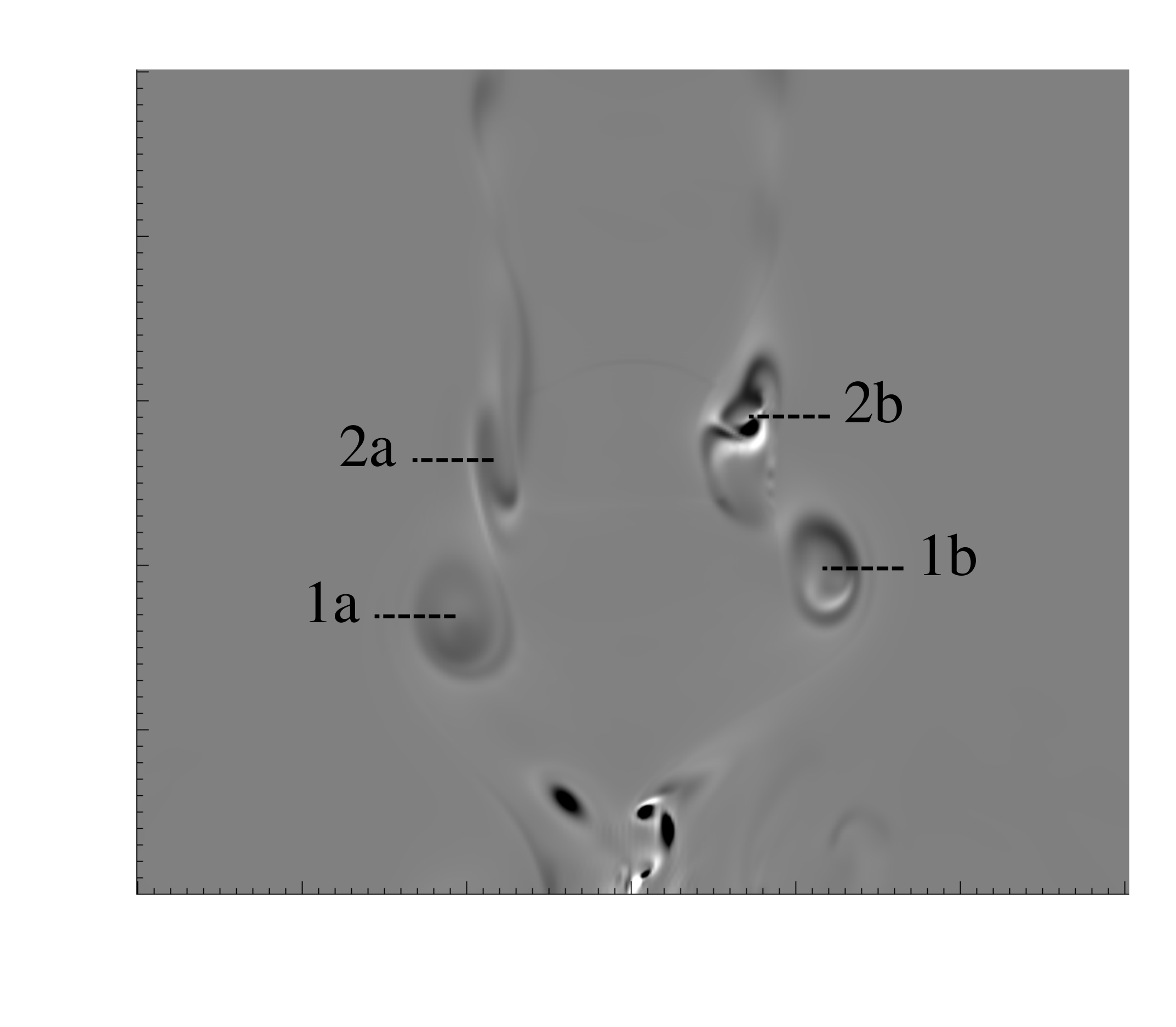}}
\subfloat[]{\includegraphics[width=0.33\textwidth]{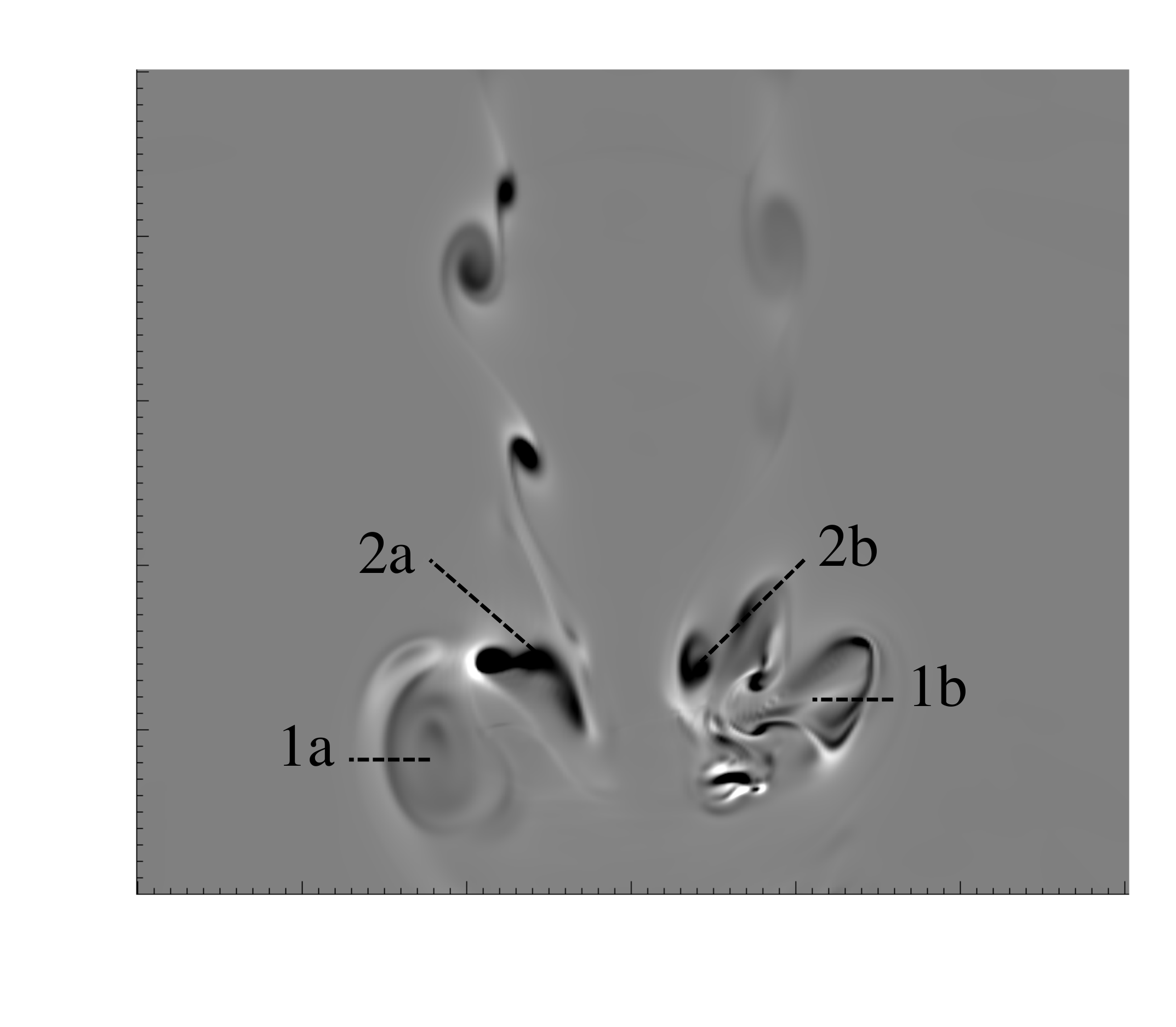}}\\[-9mm]

\subfloat[]{\includegraphics[width=0.33\textwidth]{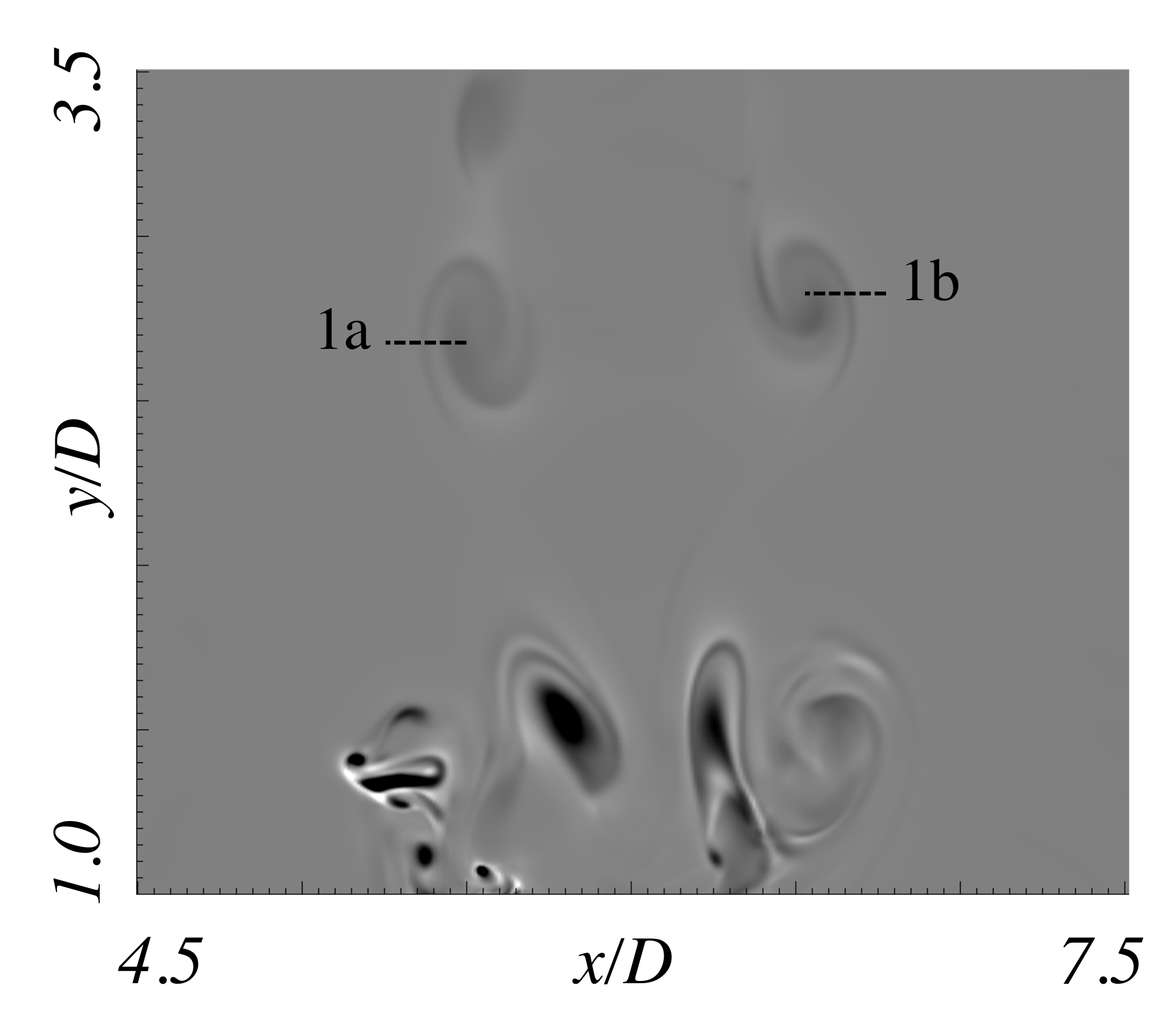}}
\subfloat[]{\includegraphics[width=0.33\textwidth]{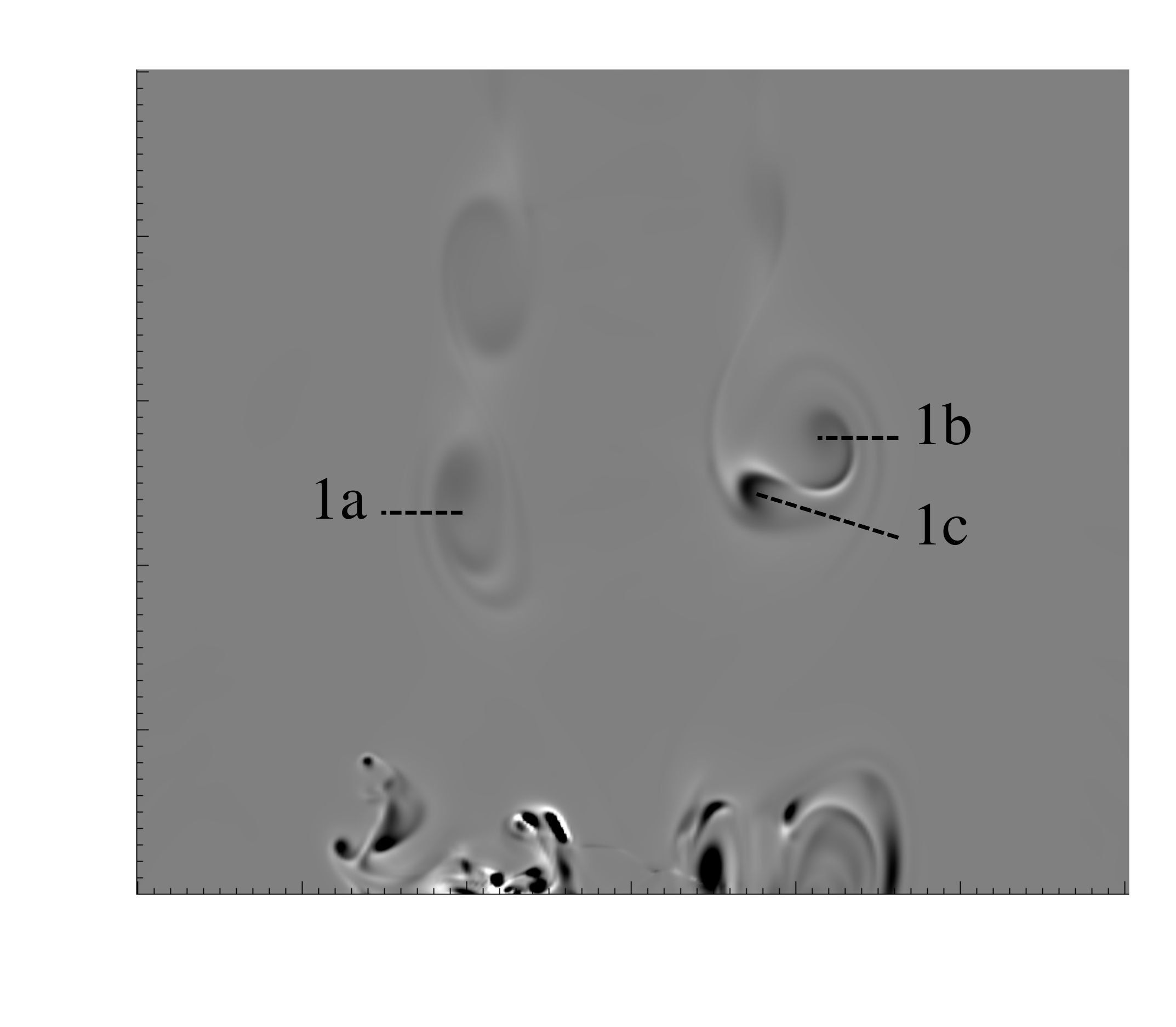}}
\subfloat[]{\includegraphics[width=0.33\textwidth]{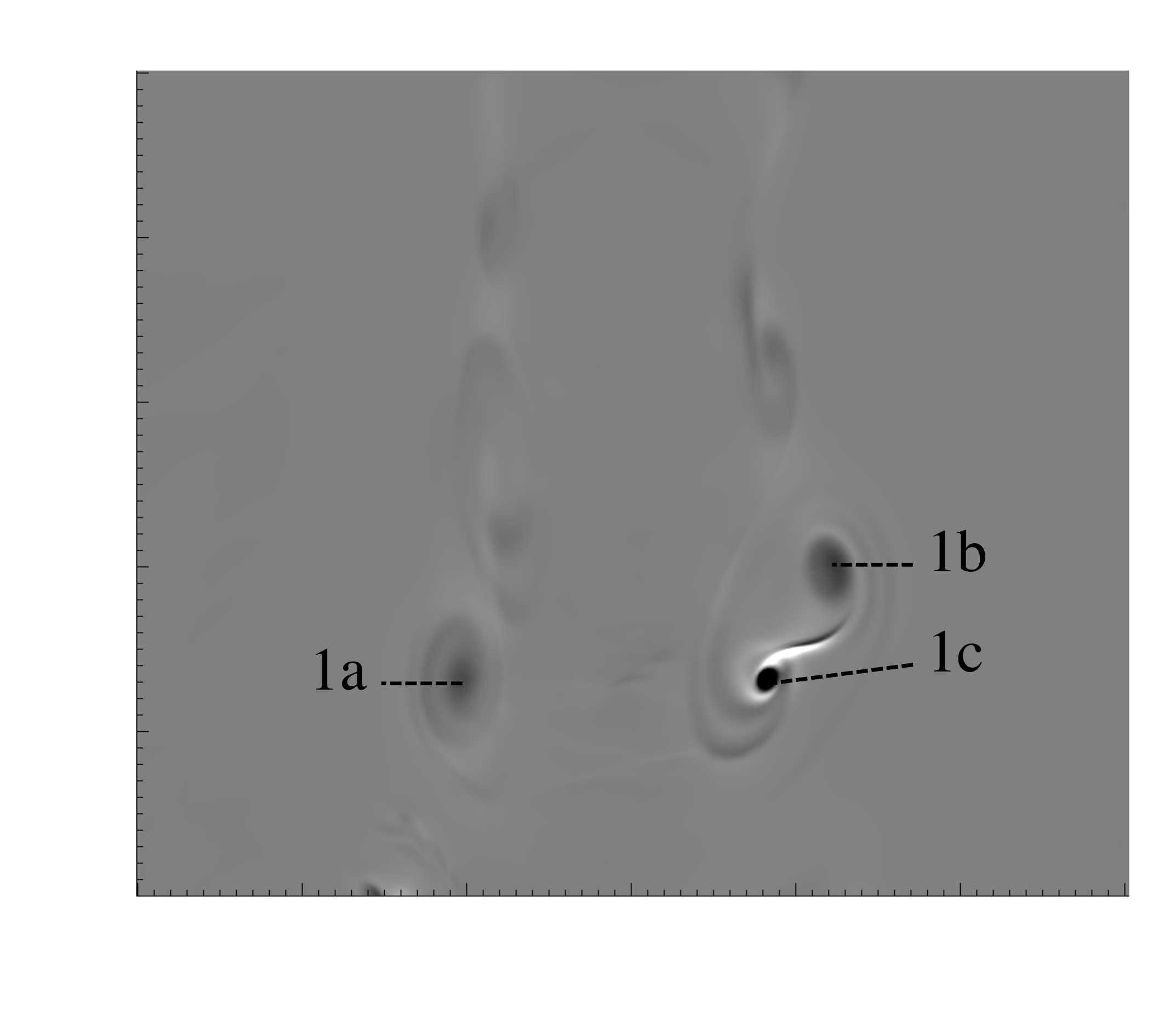}}
\caption{Leapfrogging (first row) and vortex split off (second row) in the shear layer of the free jet region in an supersonic impinging jet with $Re=8000$. The consecutive snapshots advance from the left to the right in time. Shown is $Q D/u_{\infty}^2$ [-] in the range $-85$ (white) to $85$ (black).}
\label{fig:leapfrog}
\end{figure}

Vortex rings develop axisymmetric in shear layer of the free jet region due to a Kelvin-Helmholtz instability. Travelling downstream they grow and tend to develop an asymmetry. However, this asymmetry is due to the influence of the acoustic field. The mode of the impinging jet is toroidal, no flapping and no helical mode can be observed for the investigated set of parameters. During the movement in streamwise direction two different phenomenons are observed. Vortex rings can split off a new vortex or leapfrogging can occur. Both effects can be seen in figure \ref{fig:leapfrog}. In the first row are three consecutive snapshots that show leapfrogging. Two similar vortex rings (1a,1b) and (2a,2b) are travelling downstream behind each other (left). Due to their mutual interaction the frontal one decelerates and increases its diameter whereas the rear one accelerated and shrinks in diameter \citep{Riley1998}, (middle). In the next step (right) the rear vortex ring (2a,2b) passes through the front ring (1a,1b). Depending on the positions of the vortices , the process can either be complete, before the vortices interact with the shock or both events coincide. In the second row of figure \ref{fig:leapfrog} the split off of a vortex is shown. While the vortex ring on the left side (1a) is unchanged, the ring splits on the right side (1b,1c). Thereby the new developed part (1c) takes the position of the original structure (1b), which moves out of the free jet region and slows down.

\subsection{Influence of the Reynolds number and the ambient temperature}
\label{sec:cycle3300}

The mode described above is based on the computation (\#3) of a supersonic impinging jet with a Reynolds number of 8000 and an ambient temperature of $293.15$ K, which is equal to the total inlet temperature. Two more simulations were carried out. The first one (\#2) has the same temperatures, but a lower Reynolds number of 3300. Analysing the data, we observe exactly the same mechanisms and feedback loop as in the case of $Re=8000$.

 In the second simulation (\#1) at $Re=3300$ also the ambient and wall temperature were changed to $T_{\infty}=T_W=373.15$ K. The total inlet temperature was not changed, so we have a cold jet in a hot environment, which is typical for cooling configurations. This simulation shows a specific characteristic. The flow changes between two modes, which have the same frequency $Sr\approx0.35$. This effect occurs also in free jets, as described in section \ref{sec:jet_instab}. All calculated frequencies are summarised in table \ref{tab_tones}. The modes of this specific simulation are denoted A and B, but are different from the labels A and B of the free jet screech, as described in section \ref{sec:jet-modes}.
 
In mode A strongly axisymmetrical vortex rings develop with the characteristic frequency. Those vortex rings are so far from each other that they do not interact and no leapfrogging can be observed. Each vortex ring behaves exactly like the head vortex described in section \ref{sec:DMD}. The dynamic mode decomposition of this mode is described in \cite{WilkeSesterhenn2016}. Due to the strong symmetric flow field, the stagnation point is not disturbed as strong as in simulation \#3. As a result, no high pressure lumps are developed and therefore only one standoff shock is created for each period for this mode. This one is created at $y/D\approx0.25$, moves and interacts with the (only) vortex ring at $y/D\approx0.75$.
 
Mode B by contrast is equal to the mode observed in the simulations \#2 und \#3. Here the vortex rings develop more frequent and allow leapfrogging. Anyway, in both modes the head vortex formates with the same frequency. 

If one ore more sound waves are produced within each cycle does not affect the frequency of the impinging tone, since the frequency of the head vortex and so the frequency of the entire cycle does not change. This leads to no difference in the emanated sound, as shown in section \ref{sec:sound}.

\begin{figure}
\captionsetup[subfigure]{labelformat=empty}
\centering
\subfloat[Mode A]{\includegraphics[width=0.33\textwidth]{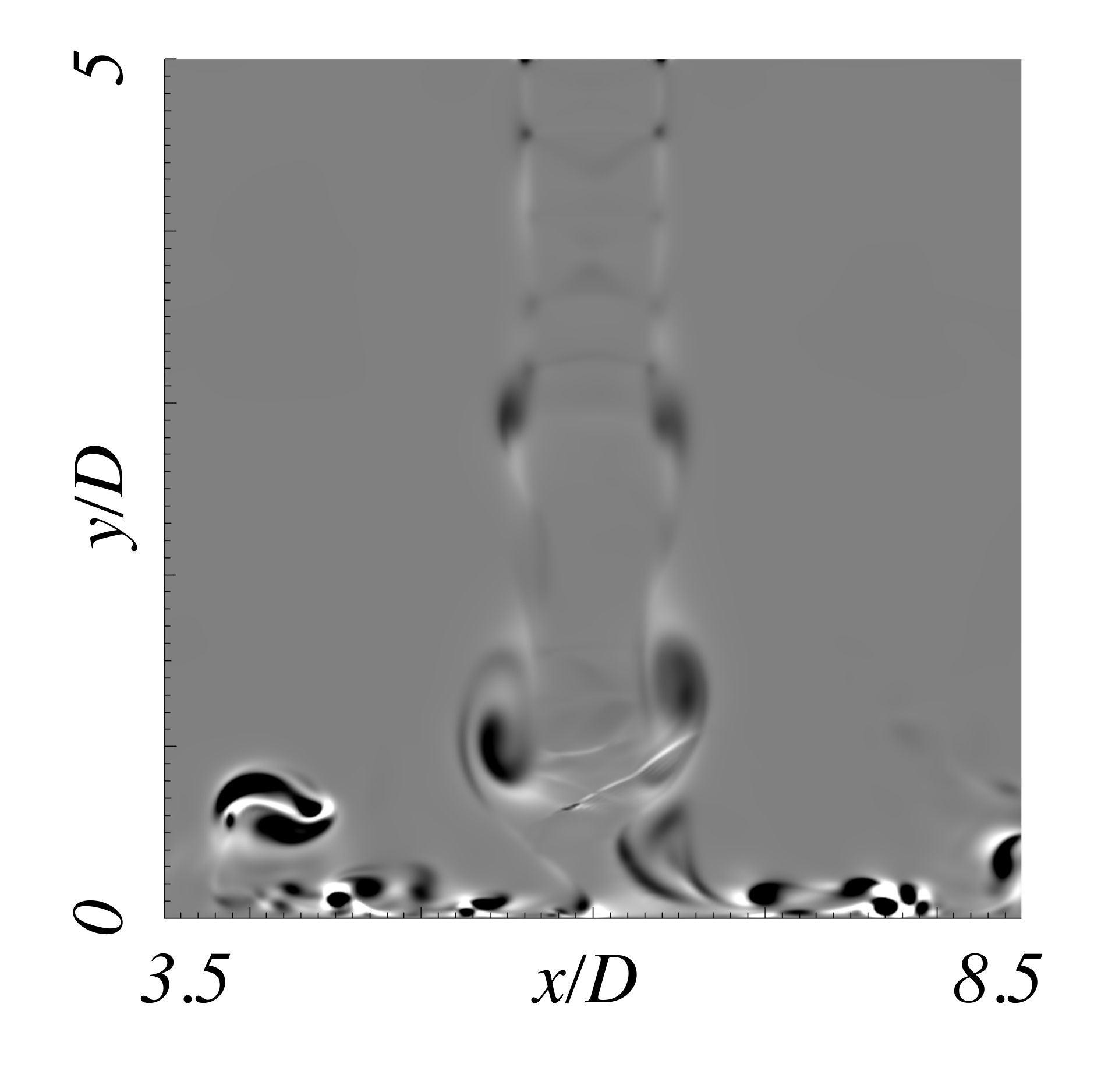}}
\subfloat[Mode B]{\includegraphics[width=0.33\textwidth]{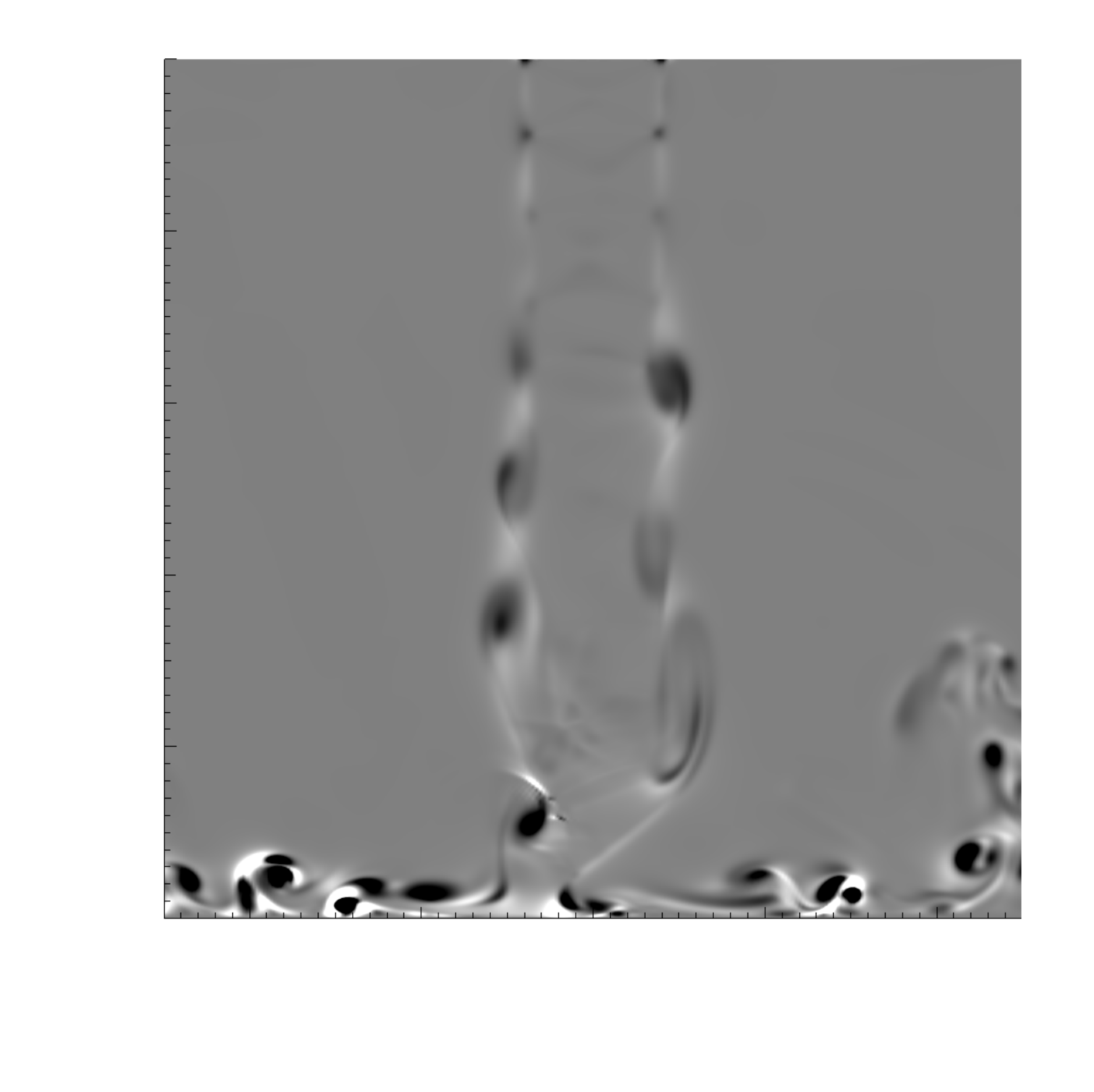}}

\caption{Two different modes exist for a cold impinging jet at $Re=3300$ ($T_0=293.15$ K) in a hot environment ($T_{\infty}=T_W=373.15$ K) with the same frequency $Sr\approx0.35$. Shown is $Q D/u_{\infty}^2$ [-] in the range $-8.5$ (white) to $8.5$ (black).}
\label{fig:3300AB}
\end{figure}

\section{Sound source mechanism}
\label{sec:sound_source_mechan}

%
%
%

\subsection{Type 1: Shock-vortex-interaction}
\label{sec:sv}

This kind of sound-emitting interaction requires two components: One shock and one vortex or an aggregation of vortices. The computational results show that multiple shocks can occur near by the stagnation point. Usually two or three shocks are simultaneously present. The system of the shocks is highly unsteady within a periodical cycle. The cycle is described in section \ref{sec:DMD}.

Shock-vortex-interactions occur also in free jets, as described by Fernandez and Sesterhenn \cite{FernandezSesterhenn2015}. However, the strength of the shock due to the impinging plate is much stronger than the one in the shock-cell-system due to the under-expansion of the jet. This results in much higher sound pressure levels in the case of a present impinging plate, on which we concentrate in this paper. Therefore the term \textit{shock} refers here always to \textit{standoff-shock}.
 
\begin{figure}
\captionsetup[subfigure]{labelformat=empty}
\centering
\subfloat[]{\includegraphics[width=0.54\textwidth]{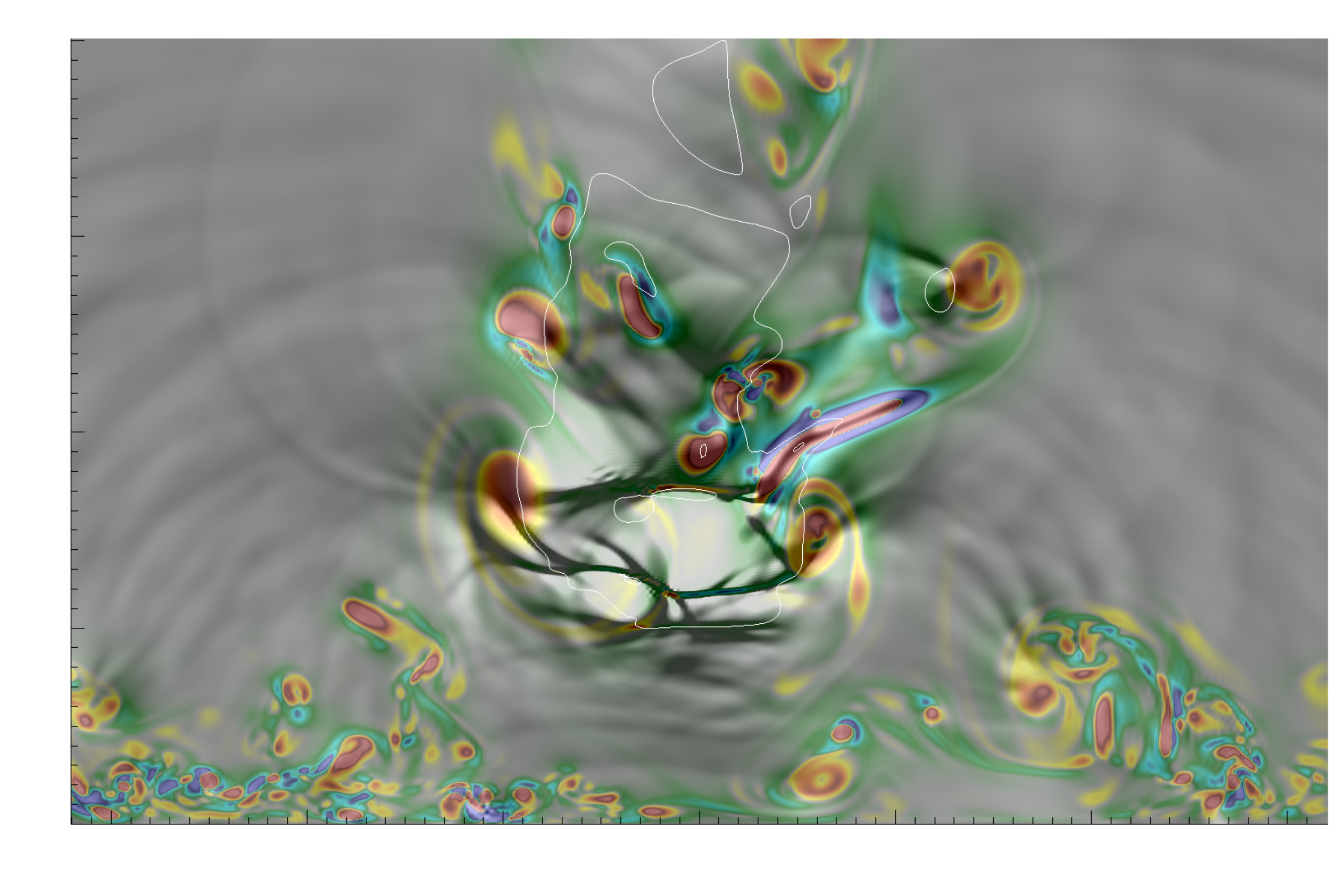}}
\hspace{0.2cm}
\subfloat[]{\includegraphics[width=0.4\textwidth]{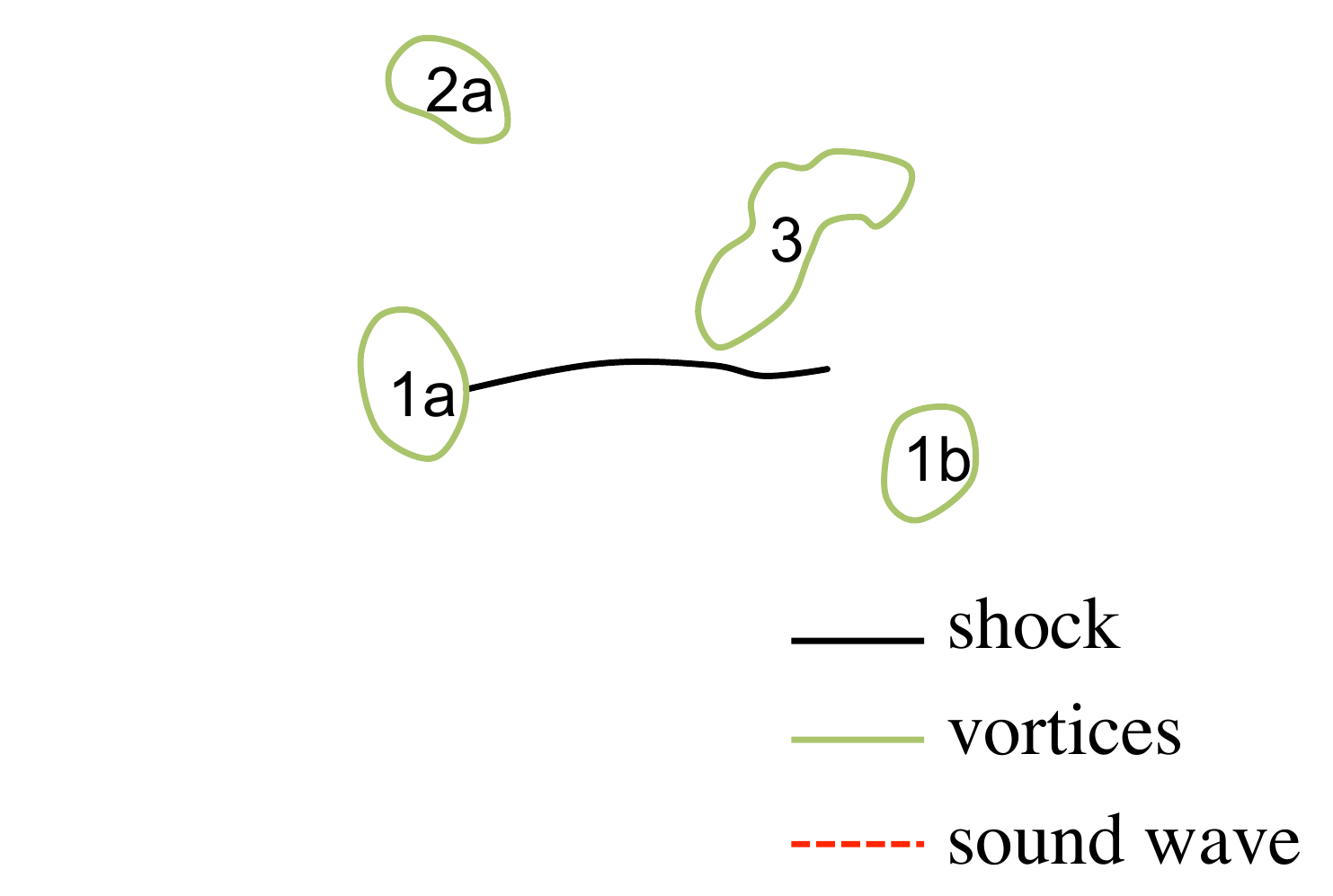}}\\[-8mm]

\subfloat[]{\includegraphics[width=0.54\textwidth]{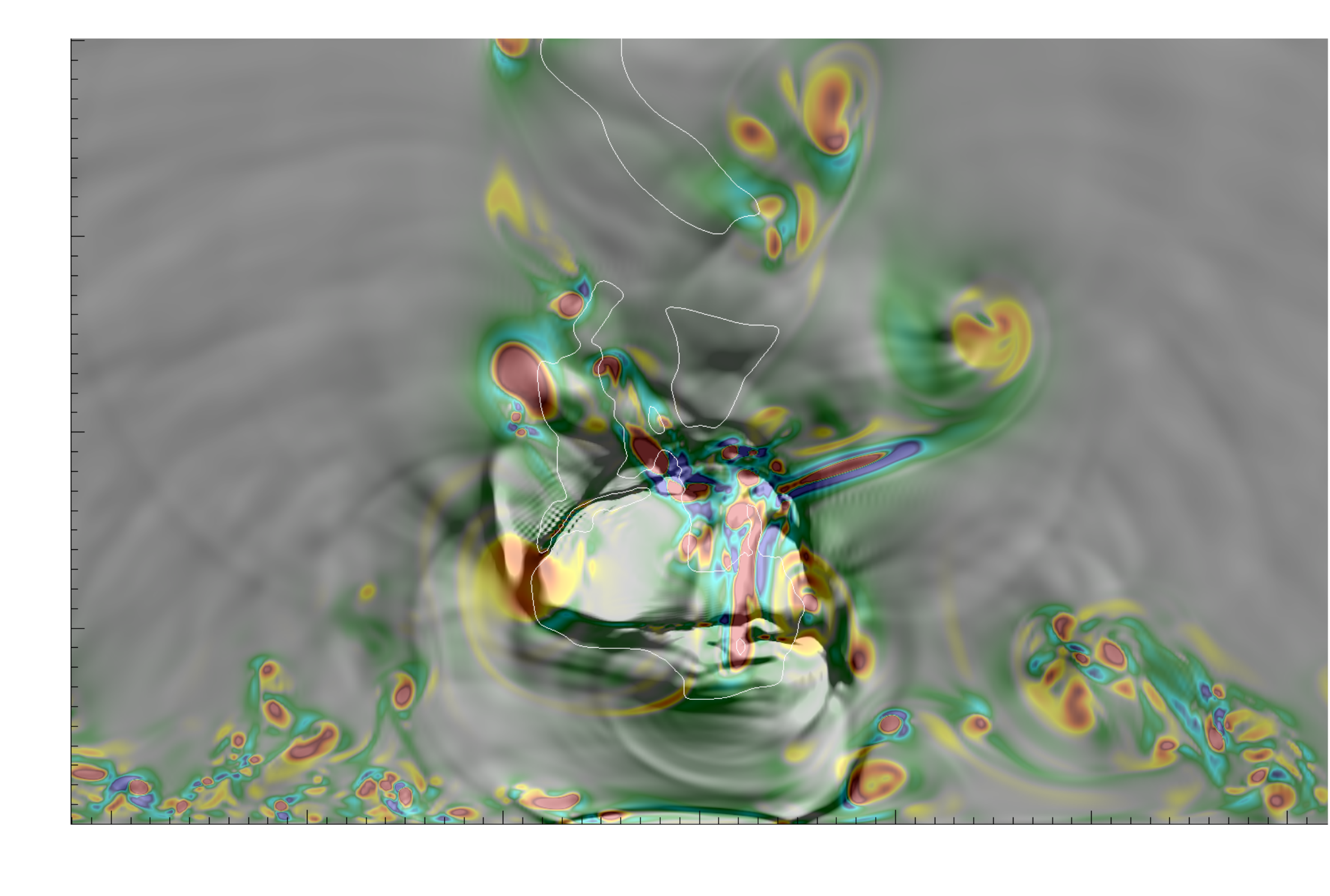}}
\hspace{0.2cm}
\subfloat[]{\includegraphics[width=0.4\textwidth]{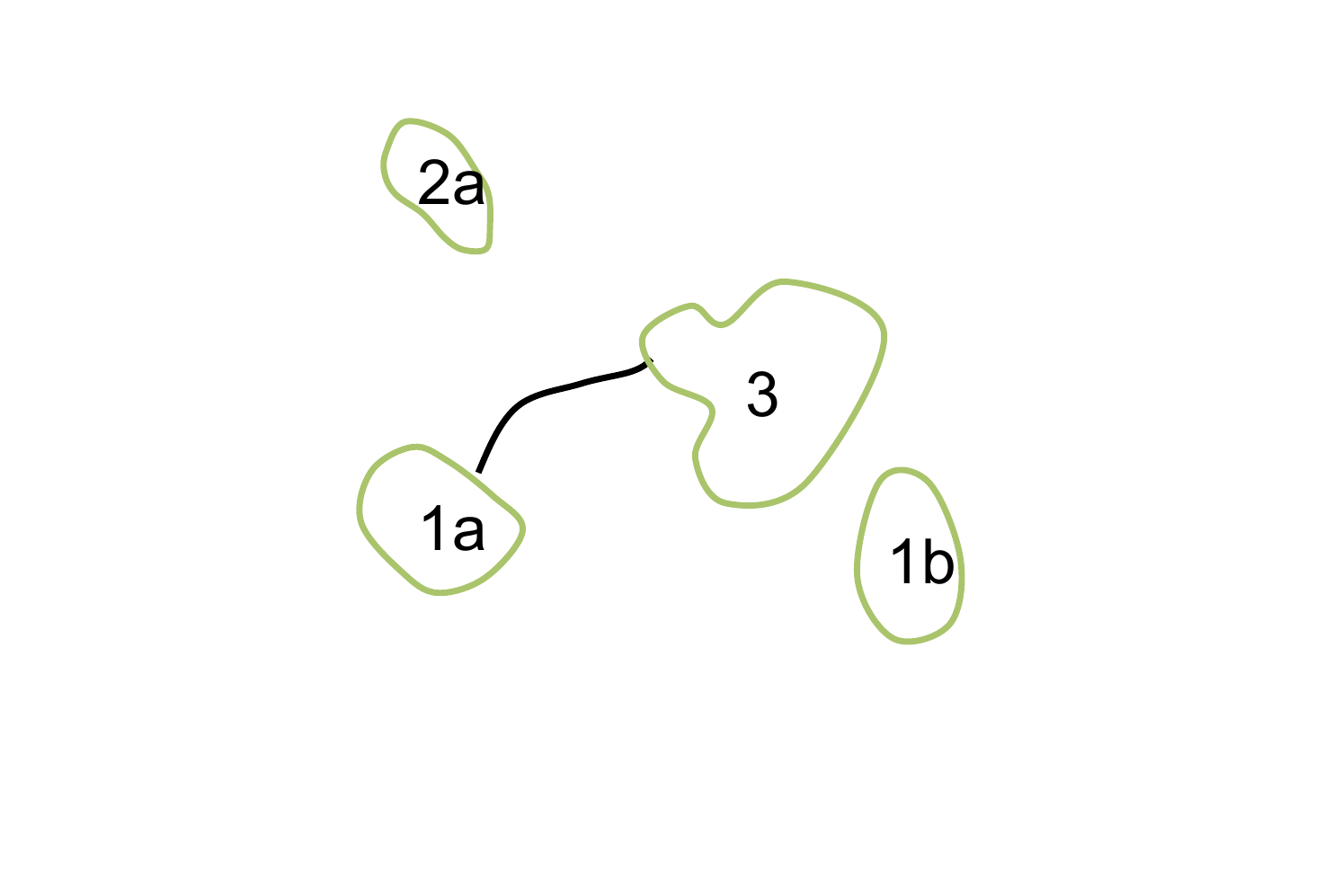}}\\[-8mm]

\subfloat[]{\includegraphics[width=0.54\textwidth]{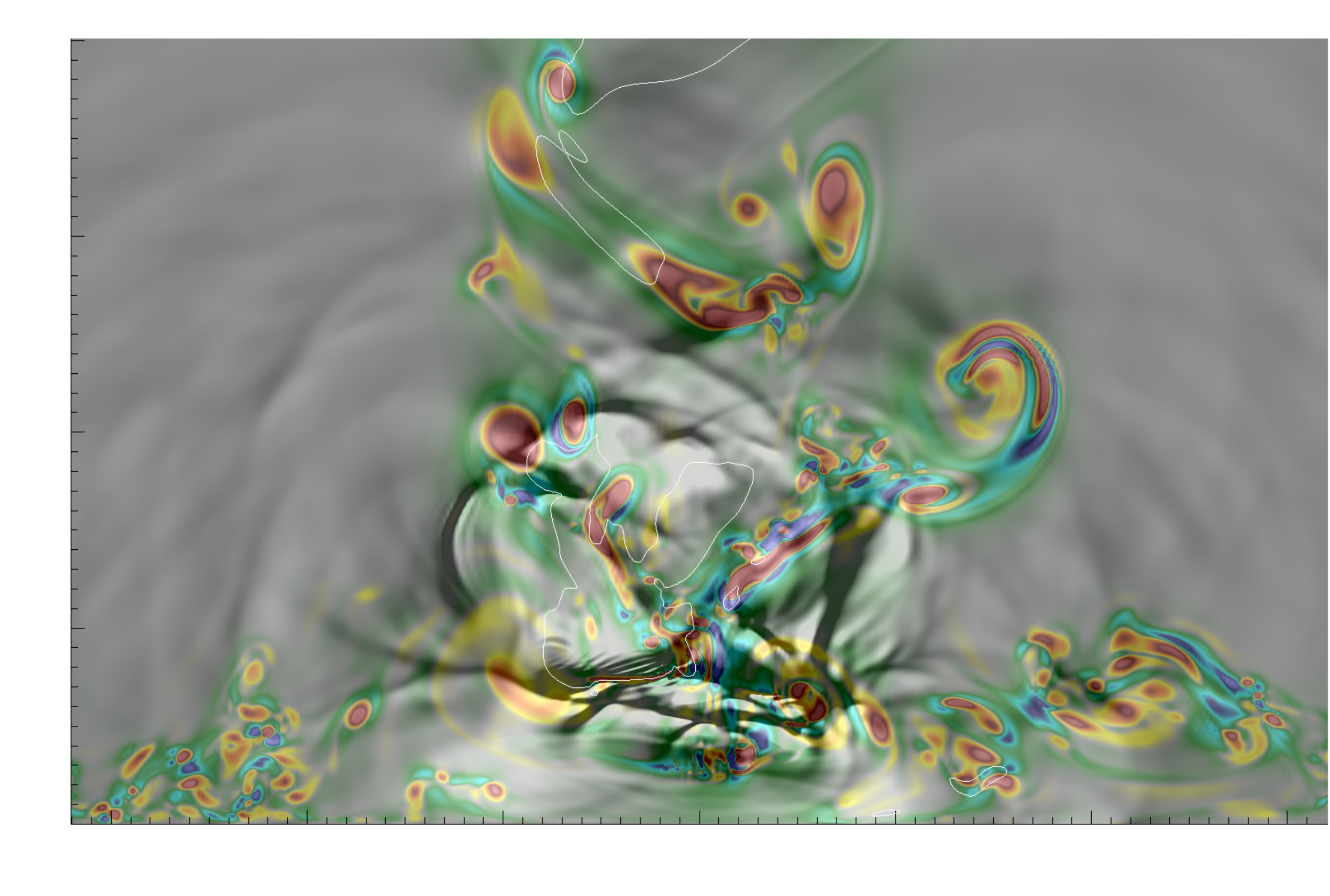}}
\hspace{0.2cm}
\subfloat[]{\includegraphics[width=0.4\textwidth]{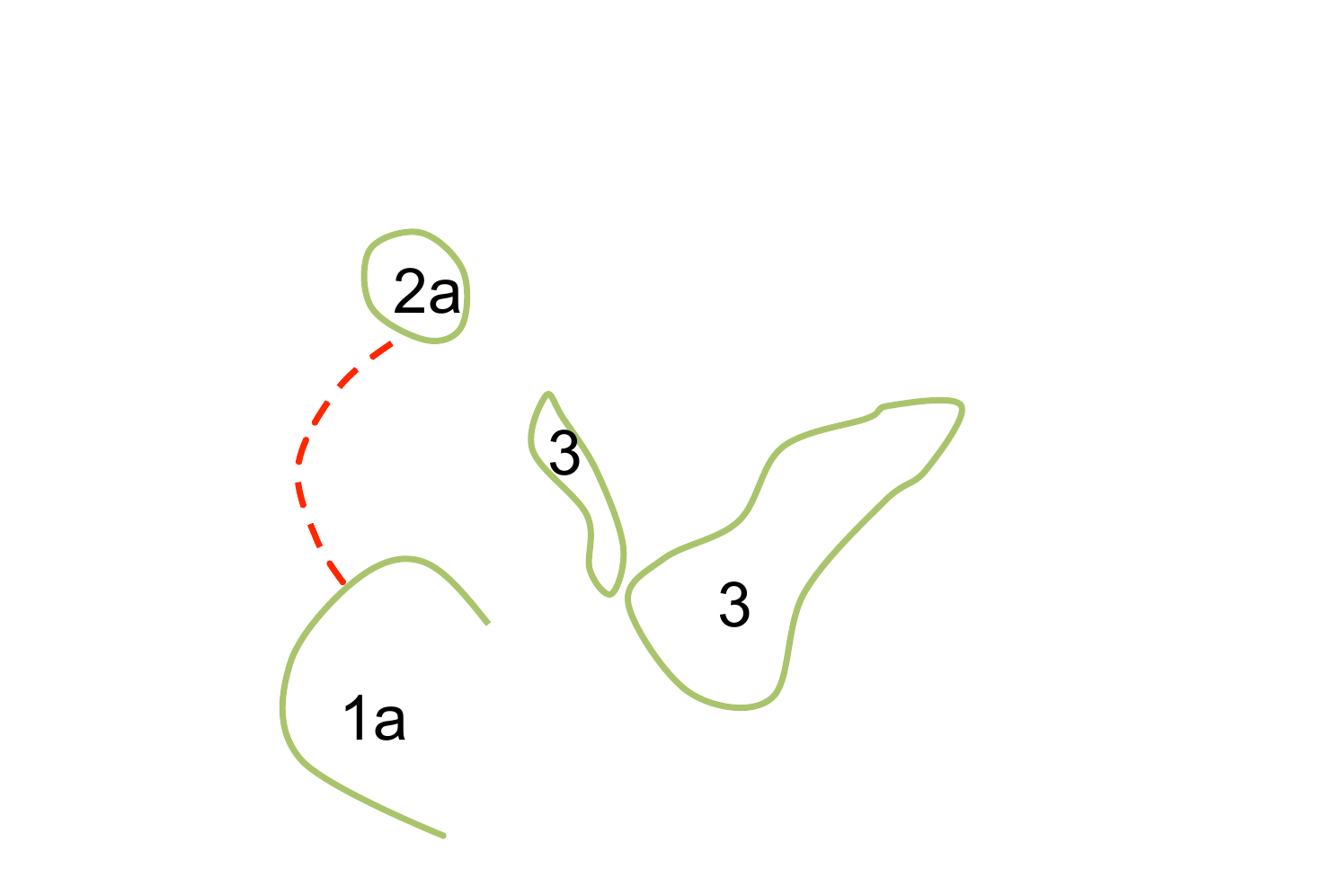}}\\[-8mm]

\subfloat[]{\includegraphics[width=0.54\textwidth]{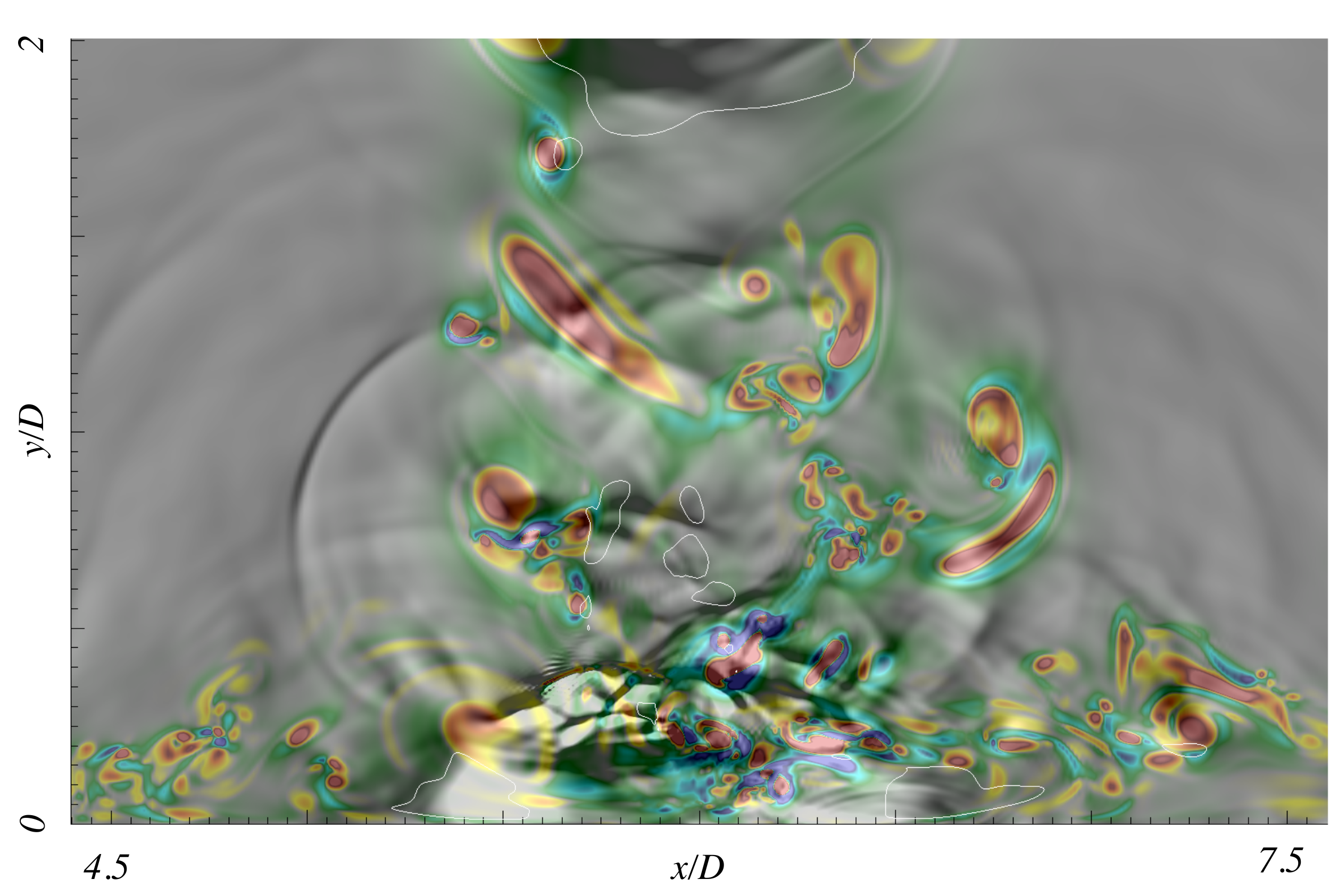}}
\hspace{0.2cm}
\subfloat[]{\includegraphics[width=0.4\textwidth]{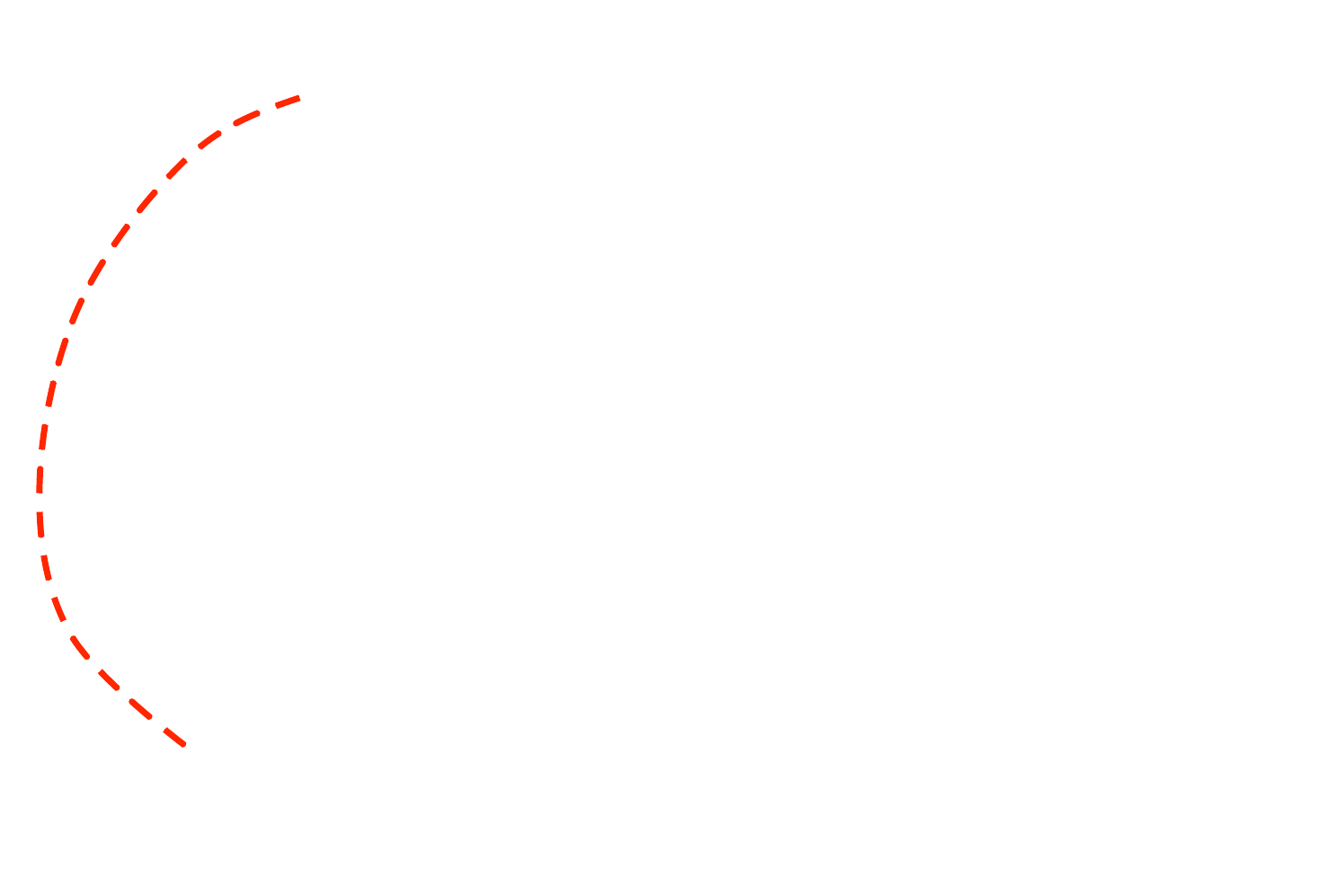}}\\[-7mm]

\subfloat[]{\includegraphics[width=0.35\textwidth]{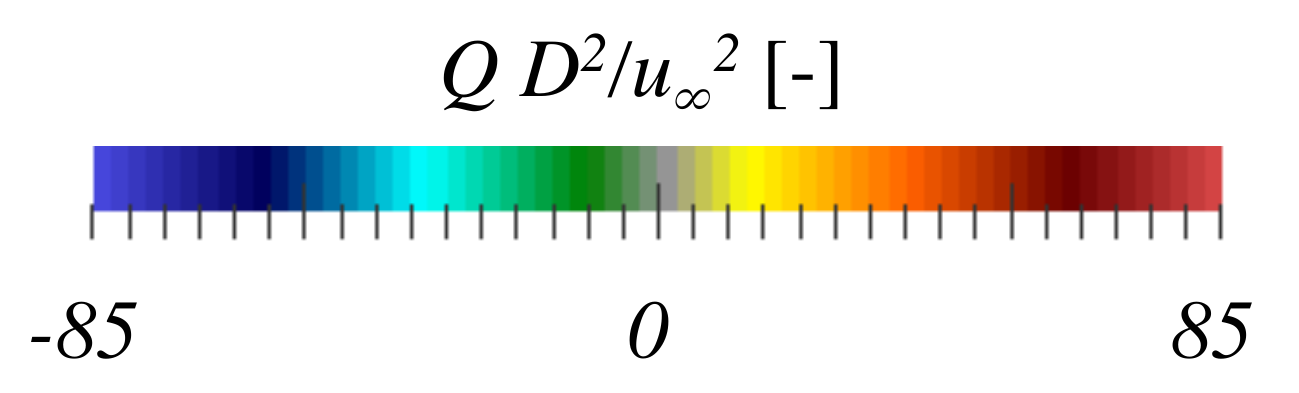}}
\hspace{0.2cm}
\subfloat[]{\includegraphics[width=0.35\textwidth]{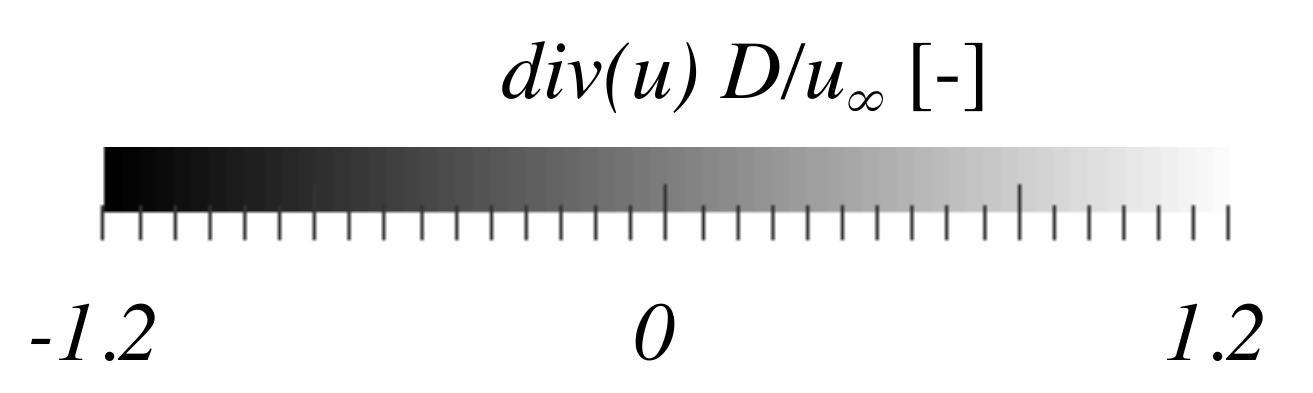}}

\caption{Shock-vortex-interaction ($Re=8000$). First column: normalised values of $Q$ and of the divergence of the velocity field $div(u)$. Second column: sketch. The snapshots (rows) are in consecutive order.}
\label{fig:sv}
\end{figure}

This sound source mechanism can involve either the main vortical structure of the impinging jet, which are the vortex rings or a vortex within a turbulent aggregation of vortices. The first case is typical for low Reynolds numbers, like $Re=3300$ and was found by Wilke and Sesterhenn \cite{WilkeSesterhenn2016}. With increasing Reynolds numbers, the phenomenon shifts to the second case. In the following, the mechanism is explained using figure \ref{fig:sv} which shows snapshots of the simulation with $Re=8000$. All snapshots are a section of a slice through the jet axis. In the first column normalised values of $Q$ and of the divergence of the velocity field $div(u)$ are shown. At the starting point (first row) three shocks are present. For this mechanism only the upper one ($y/D\approx 0.85$) plays a role. For simplicity only that one is shown in the sketch. Additionally a vortex ring (1a,1b) is present, which is slightly asymmetric. The center of the ring in the left shear layer (1a) is at the same height of shock, whereas the the center of the ring in the right side (1b) is closer to the wall. A bunch of turbulent vortices (3) is above the shock. The vortex (2a) is a fragment that is left from the next vortex ring that lost its symmetric structure due to leap-frogging. This process is explained in section \ref{sec:DMD}. At this point in time the shock keeps its position due to an equilibrium between the stagnation pressure pushing the shock up and the flow pushing the shock down to the wall. The vortices however are transported by the jet with high velocity and approach the impinging plate. The vortex ring (1a,1b) is transported in wall normal direction around the shock, without interaction. Vortex (3) on the contrary crashes into the right end of the shock. As a consequence, the shock looses its equilibrium, turns to the left and strongly accelerates. This can be seen in the second row of figure \ref{fig:sv}. At this point in time the vortex bunch (3) already cut the right end of the shock. The shock transformed into a pressure wave and is now (third row of figure \ref{fig:sv}) in between the two vortices (1a) and (3), moving in north-west direction. At this point there are two possibilities for the pressure wave. The first option is shown in the forth row of figure \ref{fig:sv}: no vortex is in the way and the pressure wave can expand without disturbance. Here, the wave can pass between vortices (1a) and (2a). In this case, the wave leaves the jet and does not trigger a feedback loop. More often is the case that there is no gap for the wave to escape and the wave interacts with another vortex, that changes the direction of the wave. In this case, the wave goes through the whole jet and triggers another instability at the nozzle lip.\\

Important for this mechanism is a flow field that is at least slightly asymmetric. At low Reynolds number ($Re=3300$), we observe a flow field that switches between a mainly symmetric and a clear asymmetric state. Also the mainly symmetric state is slightly distorted, so that one side of the vortex ring touches the shock slightly before the other side and leads to the described sound wave. Those two different states are explained in section \ref{sec:cycle3300}.\\

\subsection{Type 2: Shock-vortex-shock-interaction}
\label{sec:svs}

\begin{figure}
\captionsetup[subfigure]{labelformat=empty}
\centering
\subfloat[]{\includegraphics[width=0.54\textwidth]{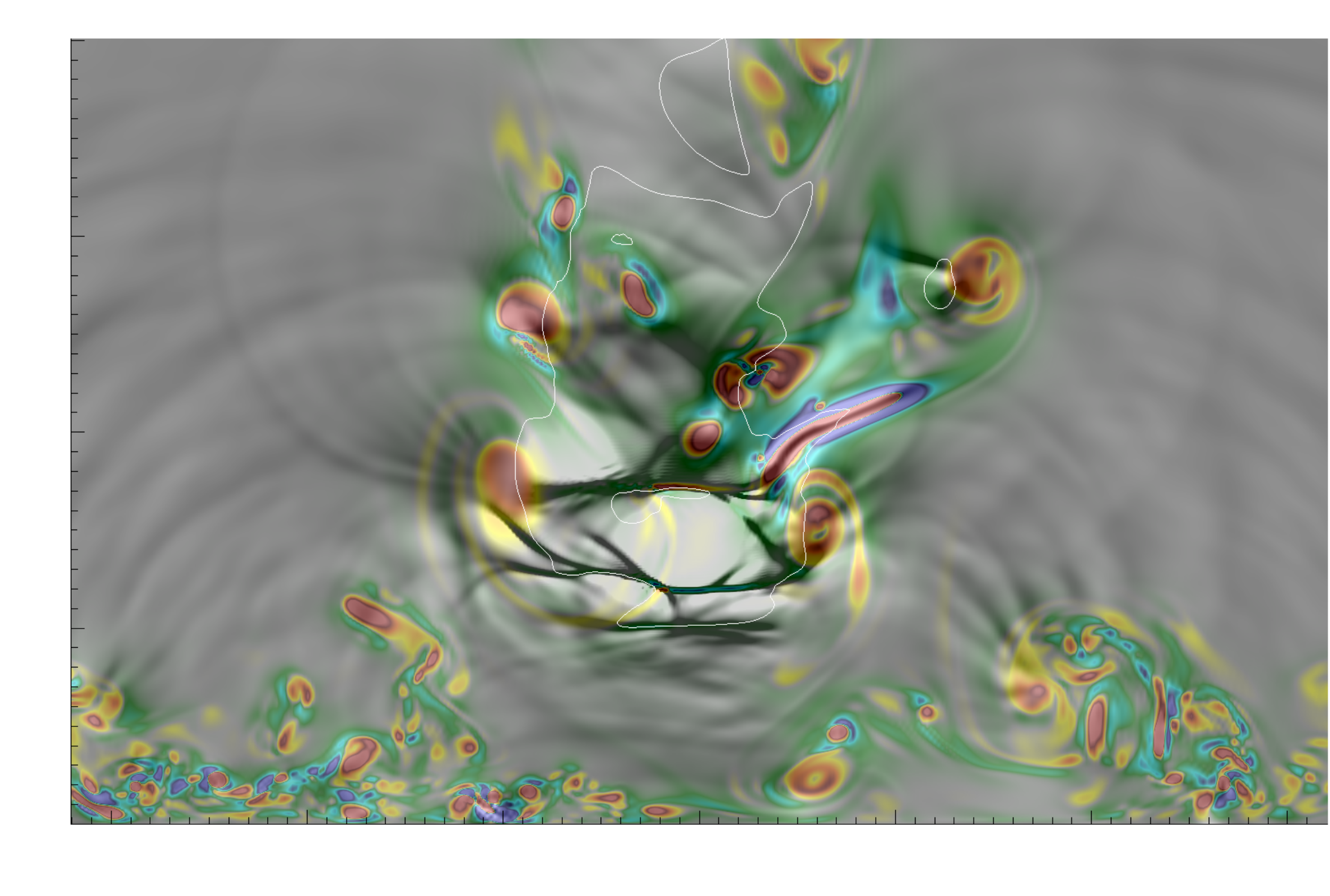}}
\hspace{0.2cm}
\subfloat[]{\includegraphics[width=0.4\textwidth]{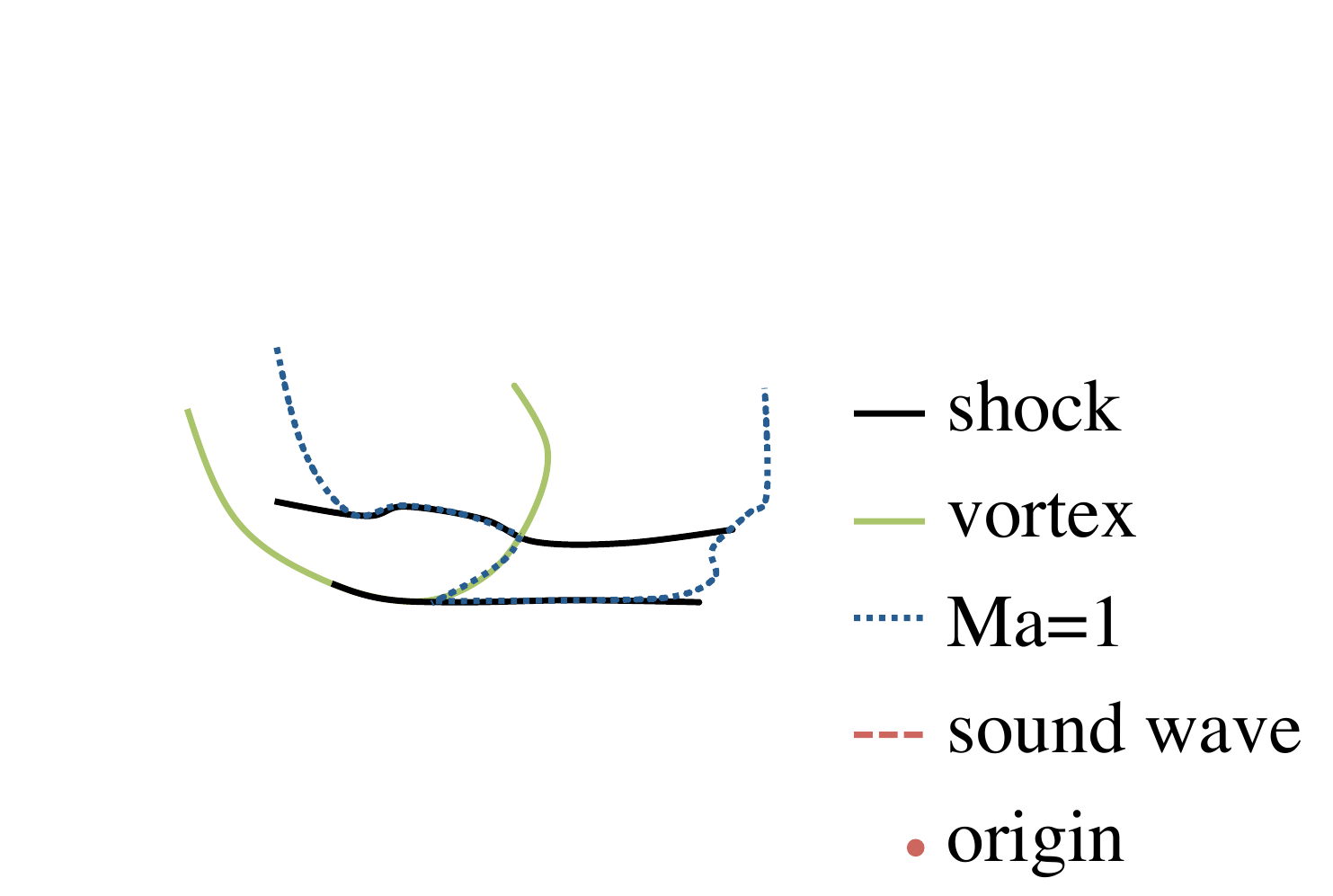}}\\[-8mm]

\subfloat[]{\includegraphics[width=0.54\textwidth]{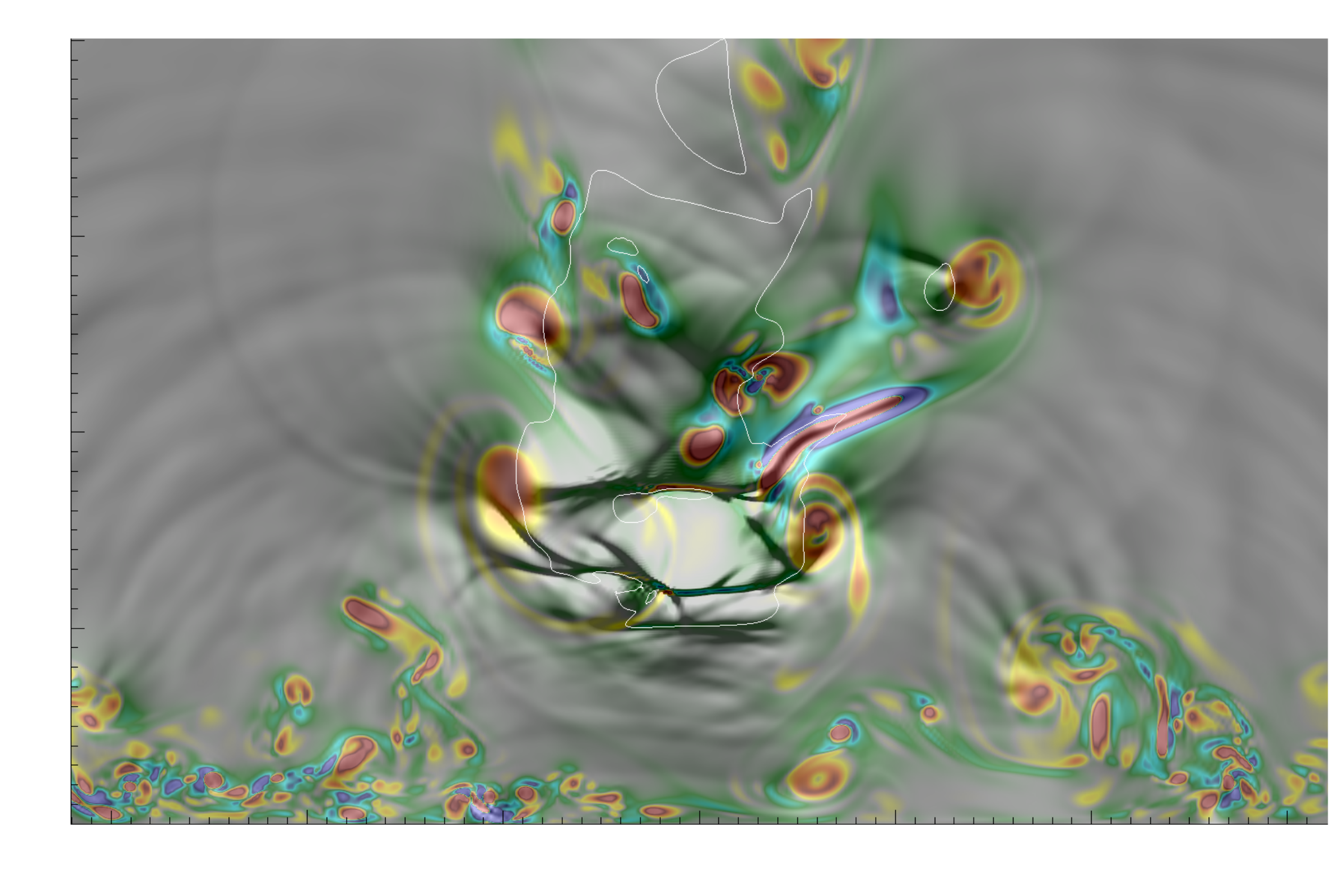}}
\hspace{0.2cm}
\subfloat[]{\includegraphics[width=0.4\textwidth]{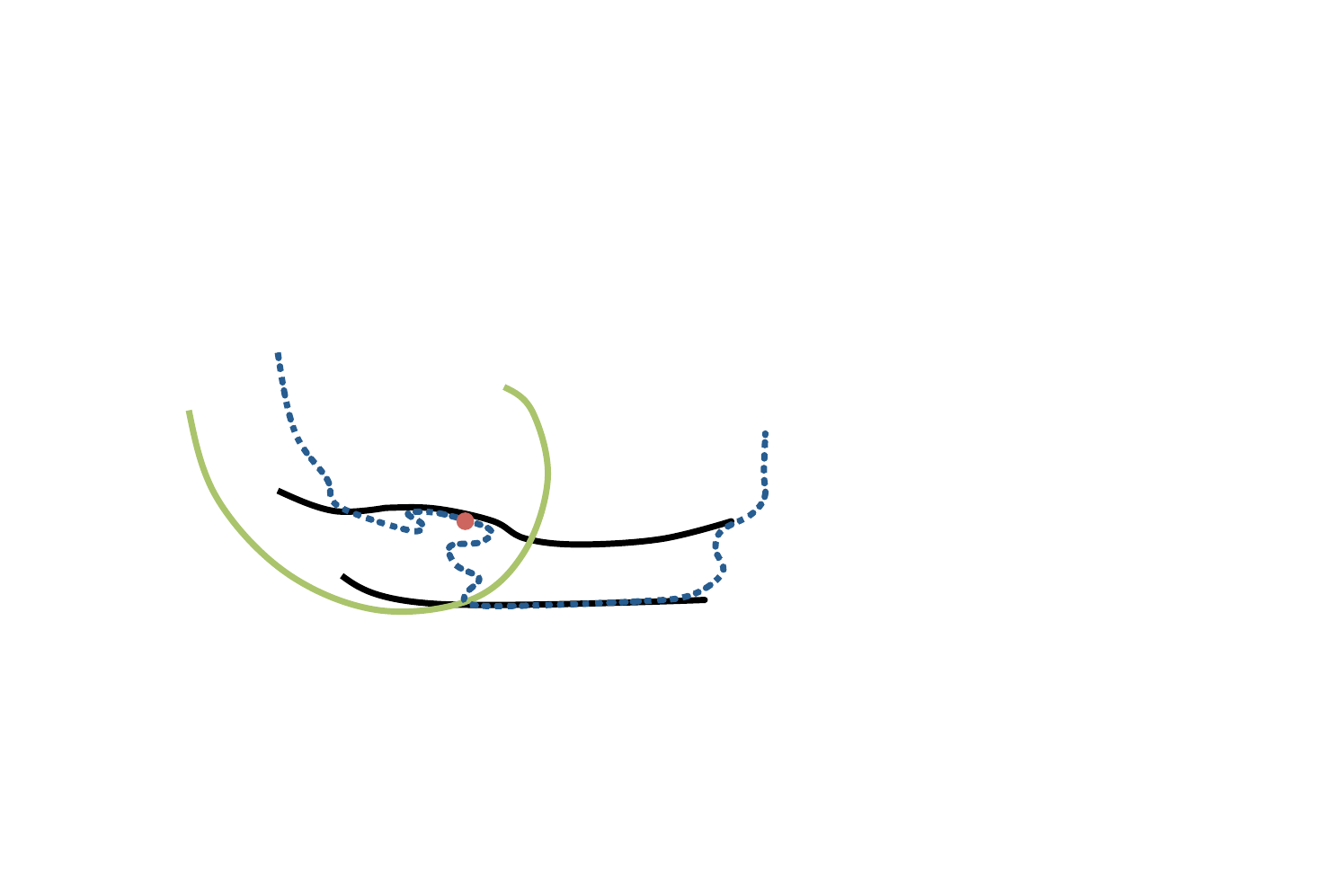}}\\[-8mm]

\subfloat[]{\includegraphics[width=0.54\textwidth]{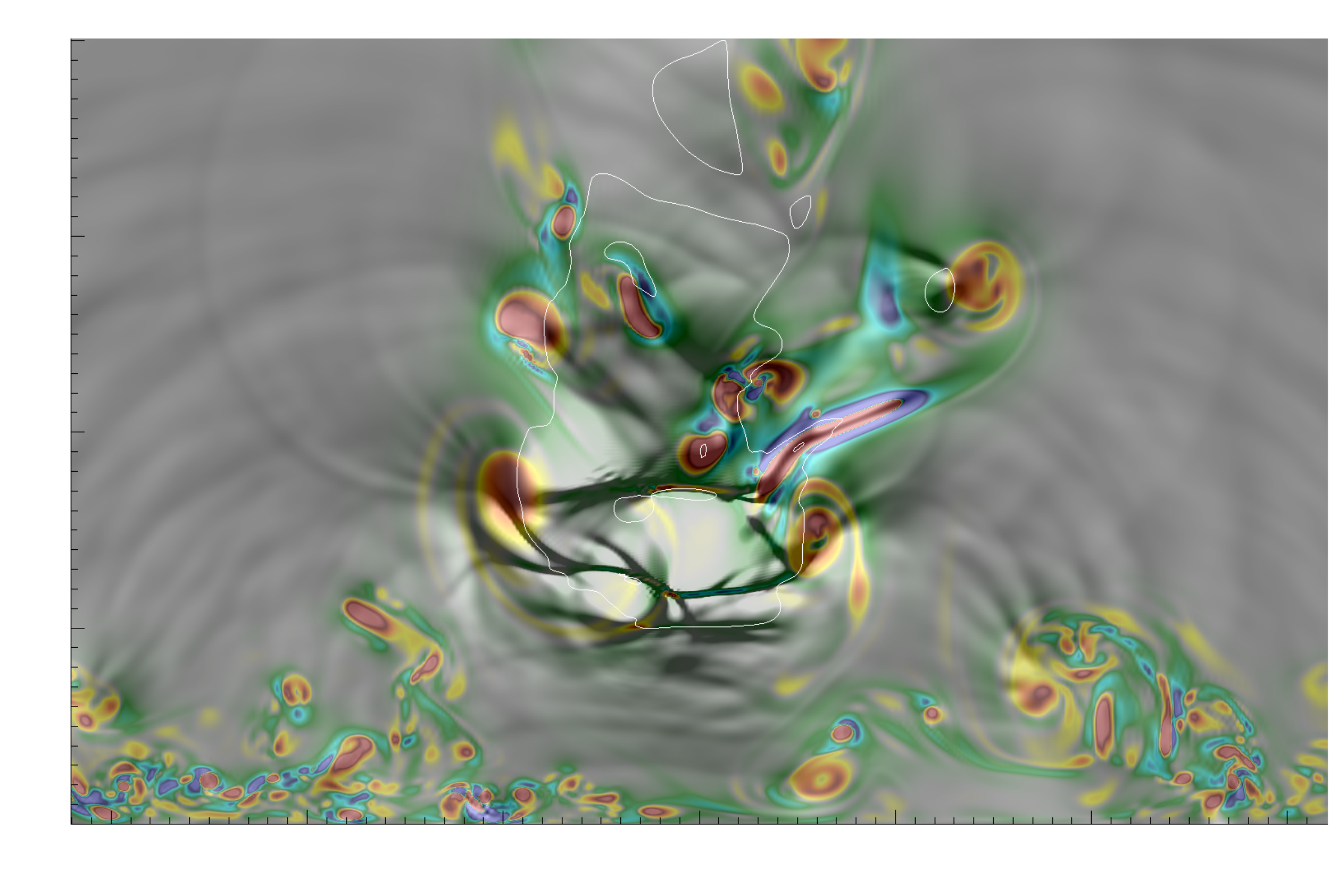}}
\hspace{0.2cm}
\subfloat[]{\includegraphics[width=0.4\textwidth]{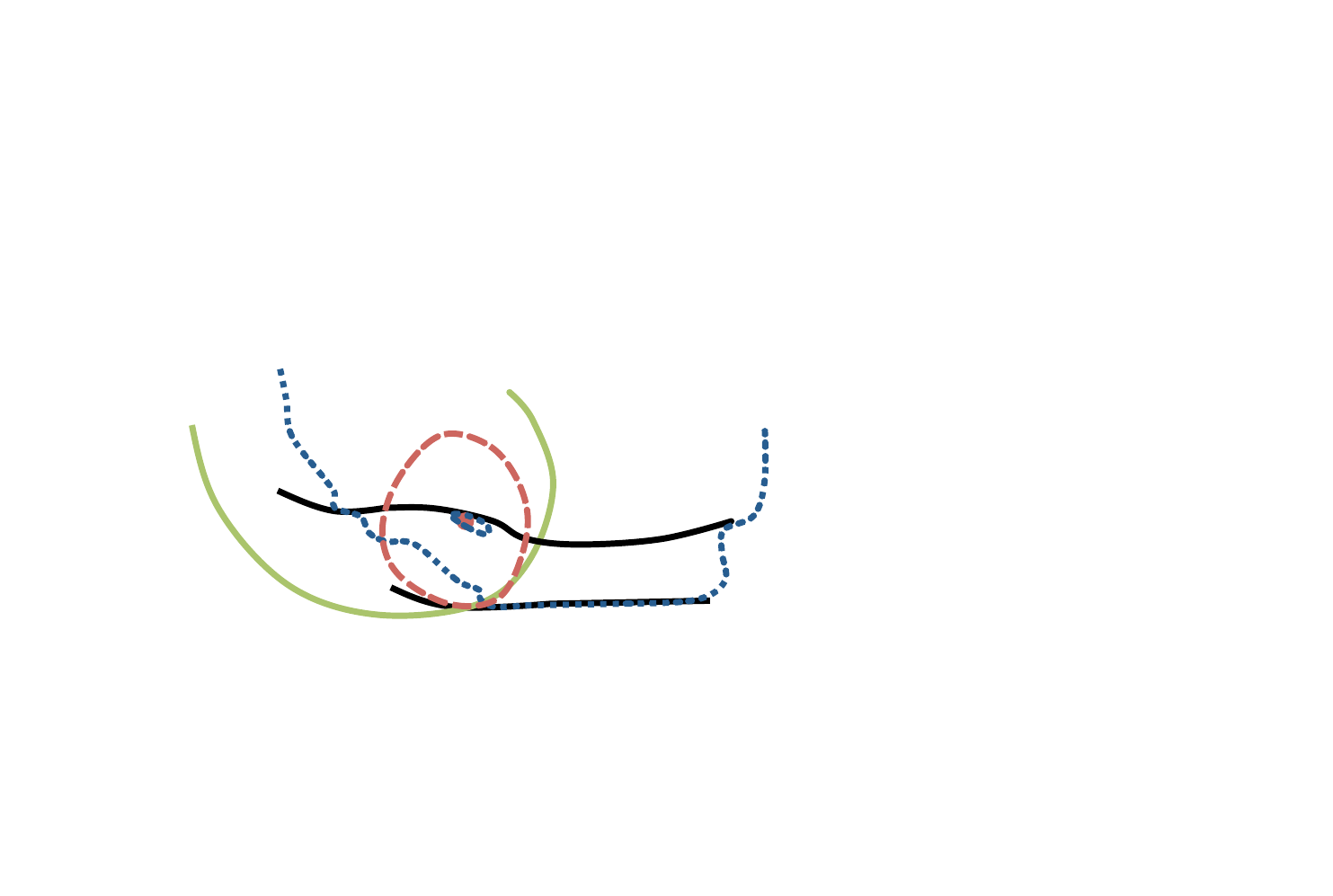}}\\[-8mm]

\subfloat[]{\includegraphics[width=0.54\textwidth]{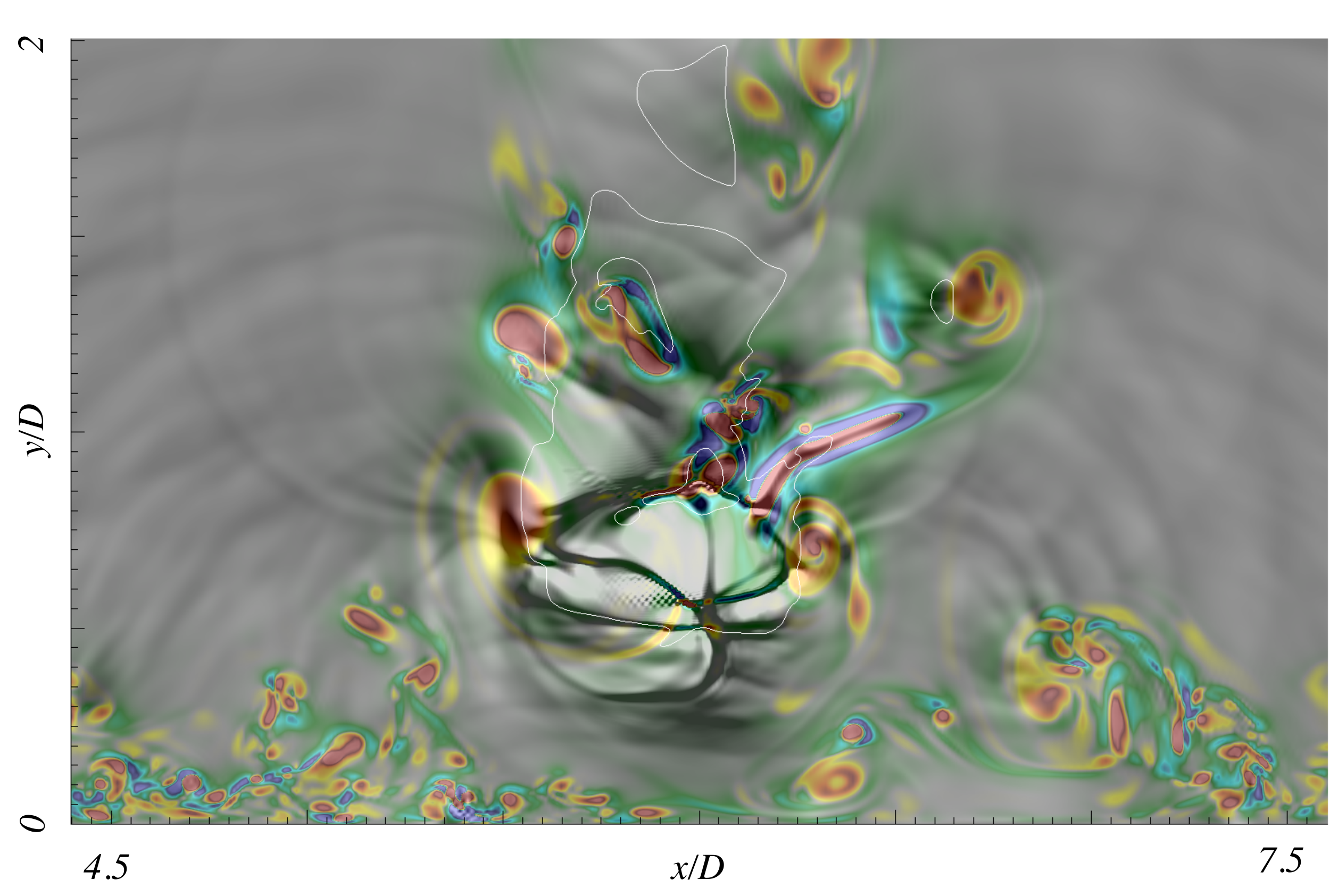}}
\hspace{0.2cm}
\subfloat[]{\includegraphics[width=0.4\textwidth]{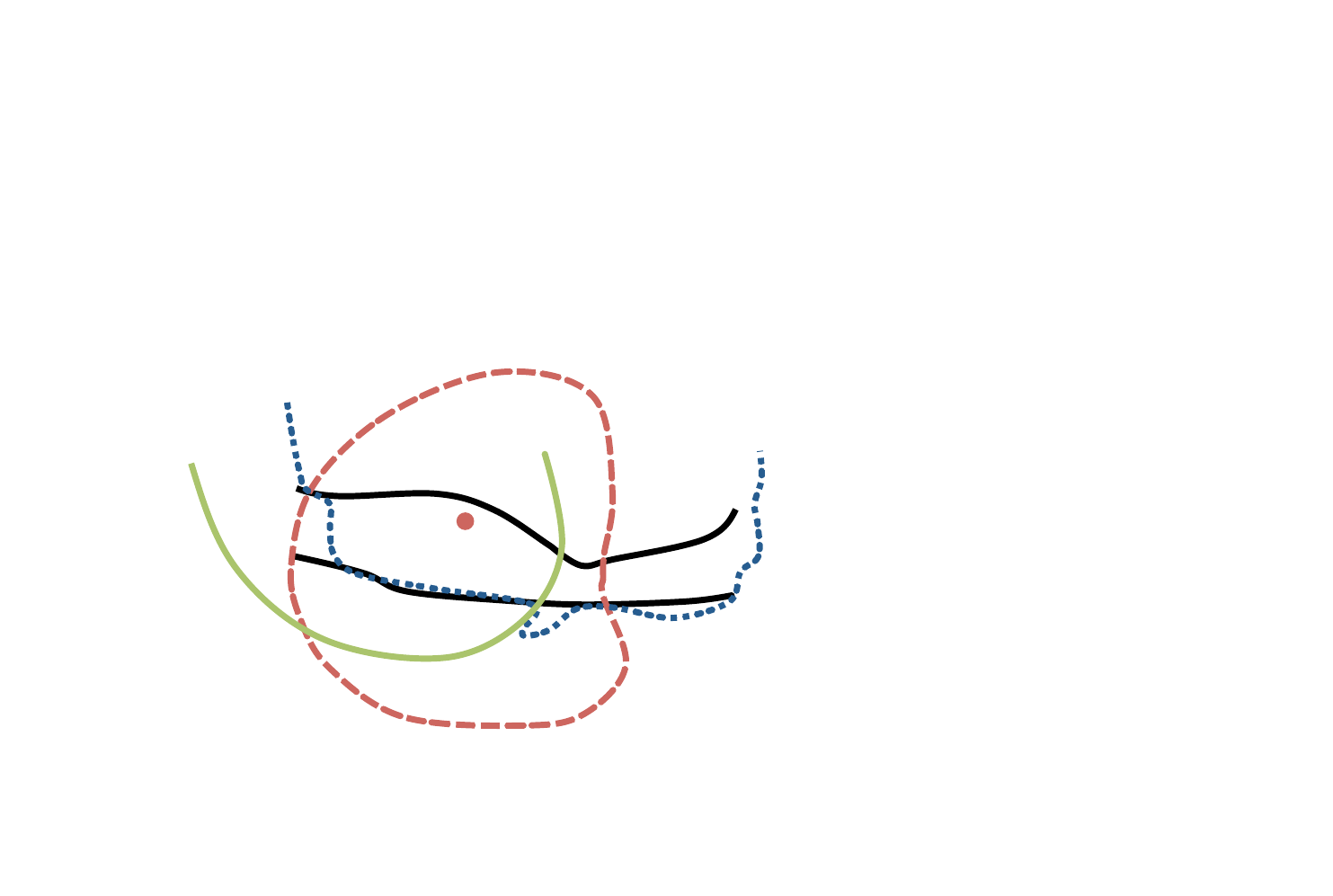}}\\[-7mm]

\subfloat[]{\includegraphics[width=0.35\textwidth]{figs/sv/color_Q.pdf}}
\hspace{0.2cm}
\subfloat[]{\includegraphics[width=0.35\textwidth]{figs/sv/color_divu.pdf}}

\caption{Shock-vortex-shock-interaction ($Re=8000$). First column: normalised values of $Q$ and of the divergence of the velocity field $div(u)$. Second column: sketch. The snapshots (rows) are in consecutive order.}
\label{fig:svs}
\end{figure}

The second kind of interaction that produces strong acoustic waves involves two shocks, a vortex ring and a sonic line. Figure \ref{fig:svs} shows snapshots of the simulation with $Re=8000$. All snapshots are a section of a slice through the jet axis. In the first column normalised values of $Q$ and of the divergence of the velocity field $div(u)$ are shown. This mechanism requires a periodical appearance and disappearance of the supersonic zone close to the stagnation point. Details about the entire cycle are given in section \ref{sec:DMD}. We start from a point in time where the supersonic zone close to the stagnation point was destroyed and a new one is transported downstream by the jet. This zone is circumscribed by the sonic line ($M=1$). As long as no obstacles are in the way, the sonic line travels together with vortex rings, but slightly ahead of them. Travelling further downstream the supersonic zone encounters zones of high pressure, which are fragments of the high pressure at the stagnation point. As mentioned, typically there are multiple of such zones. In our example, we have three of them. Each time the sonic line faces a zone of high pressure, it stops its downstream movement for a while until the jet pushes the sonic line over the shock by continuously delivering new fluid. The vortex rings travel in the shear layer, which is outside of the high pressure zone formed only in the core of the jet. Thus they are not affected by those high pressure zones. As a consequence, the vortex rings approach the sonic line and interact. This means they influence the shape of the sonic line due to its rotating velocity components. In the first row of figure \ref{fig:svs} the sonic line is confined by the shear layer of the jet in radial direction. Streamwise it consists of three parts: on the left side, the sonic line coincides with the upper shock, whereas on the right side, it coincides with the lower shock. The crossover coincides with the inner border of the left side of the vortex ring. The sound wave is produced when this arrangement collapses: The vortex is not able anymore to separate the sub- and supersonic areas. This can be seen in the following two time steps (second and third row of figure \ref{fig:svs}). The sonic line looses its connection to the vortex ring and the upper shock and jumps to the lower shock so that the upper shock gets embedded in the supersonic zone.  Thereby a subsonic area is initially embedded and then collapses. A strong spheric pressure wave expands from that point. This goes through the whole jet and reaches the nozzle. The phenomenon therefore triggers new instabilities of the shear layer and is part of a feedback mechanism.

\subsection{Closure of the feedback loop}

\begin{figure}
\centering
\includegraphics[width=\textwidth]{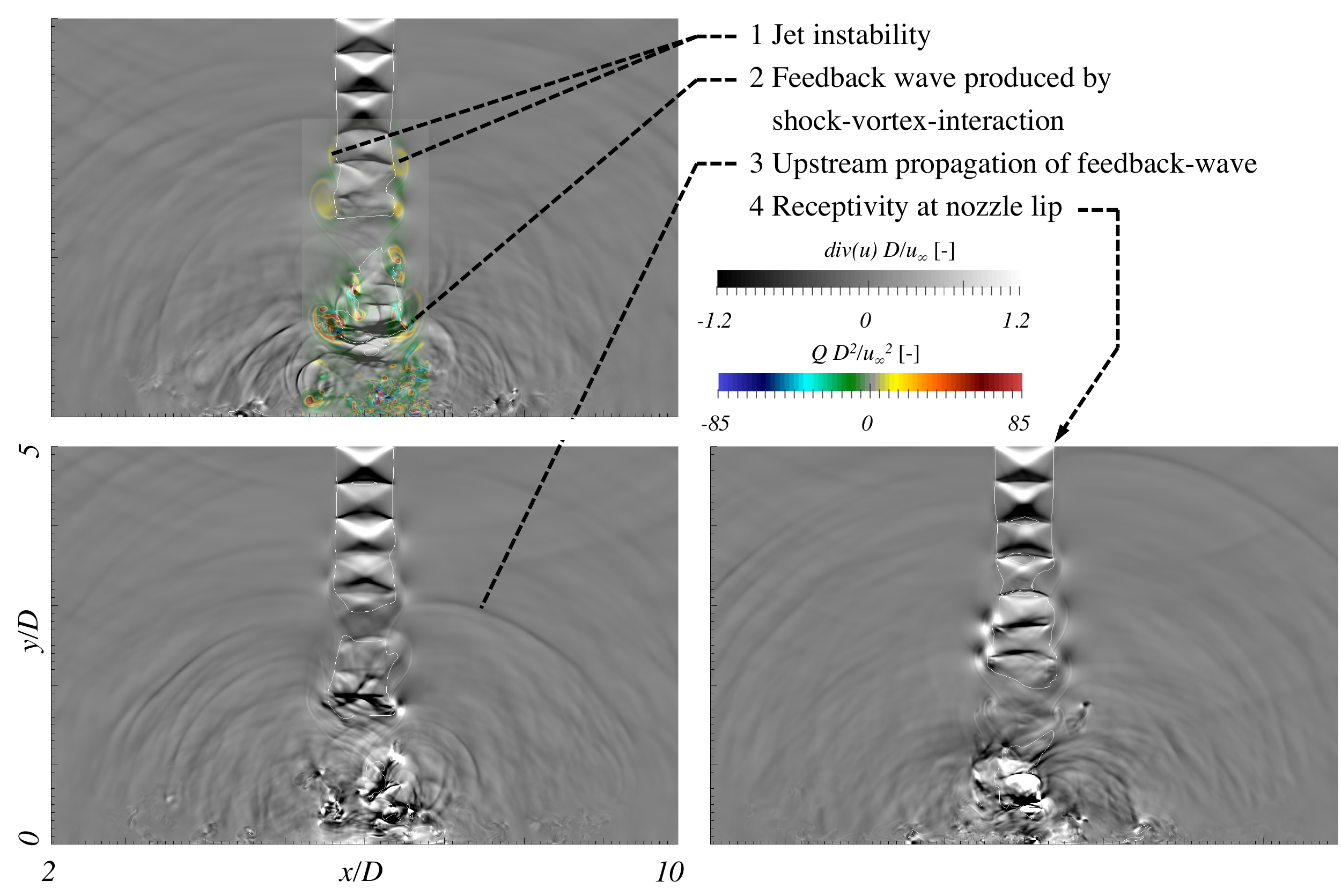}
\caption{Feedback loop of a supersonic impinging jet at $Re=8000$, inspired by the nomenclature of Raman \cite{Raman1998}, figure 1.}
\label{fig:loop}
\end{figure}

As stated in the introduction, it is generally accepted that a feeedback mechanism similar to the screech feedback loop is responsible for the impinging tones. No agreement could be found on how the loop works in detail: if the primary vortices impinging on the wall or the oscillations of the standoff shock close the feedback loop. Following the description of Raman \cite{Riley1998} of the free jet screech feedback loop, we apply the same steps for the impinging tone feedback loop:

\begin{enumerate}
	\item Jet instability
	\item Feedback wave produced by shock-vortex-interaction
	\item Upstream propagation of feedback-wave
	\item Receptivity at nozzle lip
\end{enumerate}

Vortex rings (primary vortices) develop axisymmetric in shear layer of the free jet region due to a Kelvin-Helmholtz instability (1) and perform leapfrogging as well as vortex split off's, as described in section \ref{sec:jet_instab}. Vortices interact with the standoff shocks, as described in section \ref{sec:sv} and \ref{sec:svs}, in form of shock-vortex- or shock-vortex-shock-interactions and produce strong pressure waves (2). Except for the special case, where the wave can leave the jet undisturbed, those waves usually interact again with structures of the jet and propagate as feedback-waves upstream (3). Reaching the nozzle lip, they trigger new instabilities at the shear layer (4). The feedback loop is illustrated in figure \ref{fig:loop}. The DMD showed, that it is not only one wave who triggers another wave through a direct feedback, in fact a much more complex cycle (section \ref{sec:DMD}) involving a periodical formation of head vortices and a destruction of the supersonic zone close to the stagnation point is responsible for the impinging tones.

\subsection{Emanated sound}
\label{sec:sound}

In order to obtain the sound spectra, the pressure was recorded in the near-field on three different cylinders around the jet axis at distances of two, three and four diameters. For the presented results, the position $r/D=4$ and $y/D=5$ was chosen. The upper wall has the advantage, that the velocity is zero and no flow disturbs the acoustic measurements. The choice of the radius does not influence the investigated tones (frequencies), since the different distances only move the sound pressure level up and down. For each of the 256 circumferential positions, the spectra was computed using a fast Fourier transform (FFT). The spectra were then averaged. The Strouhal number was calculated using:


\begin{equation}
	Sr=\frac{f D}{u_{\infty}} \qquad.
\end{equation}

\begin{figure}
\centering
\subfloat[Influence of Reynolds number (cold)]{\includegraphics[width=0.49\textwidth]{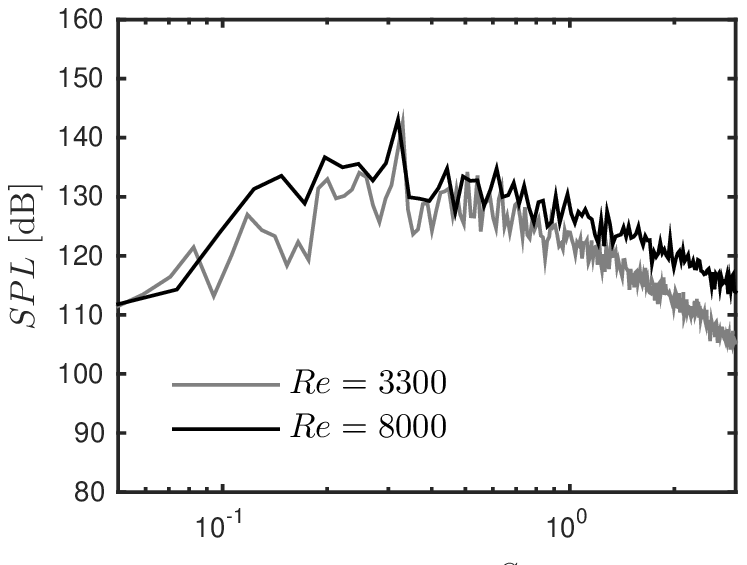}}
\subfloat[Influence of temperature at $Re=3300$]{\includegraphics[width=0.49\textwidth]{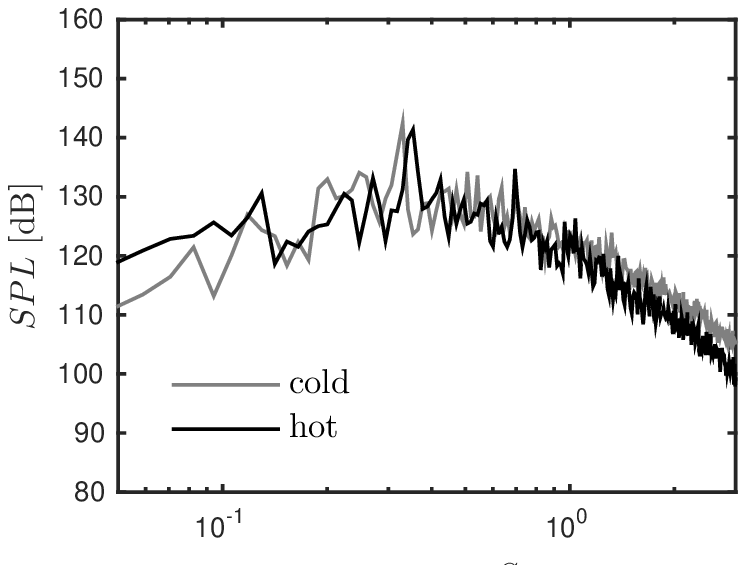}}

\caption{Sound pressure level (SPL) of different configurations of the impinging jet. Reference pressure: $p_{ref}=2 \cdot 10^{-5}$ Pa. Cold and hot classify the ambient temperature in comparison to the total inlet temperature of the jet. The parameters can be found in table \ref{tab_para}.}
\label{fig:SPLRe}
\end{figure}

$D$ is the inlet diameter, $f$ the frequency and $u_{\infty}$ the fully expanded jet velocity. In the described simulations, a hyperbolic tangent profile was used to define the inlet. This profile has a radial displacement 

\begin{equation}
	\delta_r^*=\frac{D-D^*}{2}
\end{equation}

of $\delta_r^*=0.1 \cdot D$, based on the average flow field. The effects due to the boundary layer displacement are not taken into account while computing the Strouhal number, since the diameter $D$ and not the displaced diameter $D^*$ 

\begin{equation}
	D^*= \left. \sqrt{\frac{4}{\pi} \frac{\dot{m}}{\rho v}} \quad \right|_{inlet}
	\label{eq:D_st}
\end{equation}

is used. $\dot{m}$, $\rho$ and $v$ are the mass flow, density and velocity in axial direction at the inlet. Changing the reference length to $D^*$ effects the non-dimensional frequencies.
Figure \ref{fig:SPLRe} shows the spectra for all three simulations. It can be seen (a) that the frequency of the impinging tone is nearly independent of the Reynolds number in the range of $3300 \leq Re \leq 8000$. Both simulations show a peek at $Sr=0.32$ resp. $Sr=0.33$. The frequencies of the impinging tones are summarised in table \ref{tab_tones}. The high-frequent noise increases with increasing Reynolds number.

\begin{figure}
\centering
\includegraphics[width=0.49\textwidth]{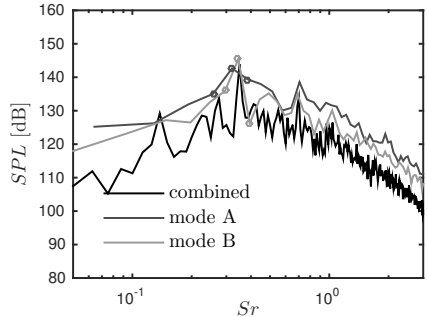}

\caption{Sound pressure levels (SPL) of the impinging jet at $Re=3300$ with a hot ambient temperature ($T_W = T_{\infty}=373.15$ K) for different time spans: Mode A and B represent the time span where the impinging jet is situated in the respective mode; combined referres to the entire pressure date (\#1), including both modes. Reference pressure: $p_{ref}=2 \cdot 10^{-5}$ Pa.}
\label{fig:SPLmodAB}
\end{figure}

In figure \ref{fig:SPLRe} (b) two impinging jets at $Re=3300$ with different wall and ambient temperatures are compared. The values of the cold respectively hot case are $T_W = T_{\infty}=293.15$ K and $T_W = T_{\infty}=373.15$ K. The total inlet temperature of the jet was kept constant at $T_0=293.15$ K. Heating the walls and therewith the ambient fluid leads to a shift of the impinging tone to slightly lower frequencies. However the profile is very similar to the cold case. This is despite the existence of an additional mode in the hot case, as described in section \ref{sec:cycle3300}. In order to compare the noise emitted by those modes, the spectra were generated additionally for the specific time span of each mode. The time spans are identical with the ones used for the dynamic mode decompositions. Figure \ref{fig:SPLmodAB} shows these spectra. It can be seen that the impinging tone is present in both cases. Furthermore the frequency is nearly identical $Sr\approx 0.35$. The small discrepancy can be explained due to the short time spans and the following coarse resolution of the Strouhal number for deeper frequencies. The data points are marked around the impinging tone. Comparing the first harmonics, we see a much smaller discrepancy due to the higher resolution of the Strouhal number on a logarithmic axis. In conclusion, the impinging tone can be either produced by only one shock-vortex-interaction per cycle (mode A) or by multiple interactions per cycle: shock-vortex-interactions and shock-vortex-shock-interactions (mode B). The frequency of the cycle is equal for both cases and is characterised by the formation of a head vortex, which is either one vortex ring or multiple vortices merged due to the leap-frogging mechanism.

\begin{table}
	\caption{Dimensionless frequencies of the impinging tones as observed in the spectra $Sr_{SPL}$ and in the dynamic mode decomposition $Sr_{DMD}$\\
	 $^*$ computed using the half frequency of the first harmonic of the tone. This is done, because the time span used for the spectrum $t_{SPL}$ is relatively short and therefore the resolution of the impinging tone frequency is coarse.}
	\begin{tabularx}{\columnwidth}{p{10mm} p{25mm} XXXXX}
	\toprule
	N$^{\circ}$ & $T_{\infty}=T_W$ [K]& $Re$ & $t_{SPL}$ [s] & $Sr_{SPL}$ & $Sr_{DMD}$\\
	\midrule
    \#1   & $373.15$ & $3300$ & $0.250$ & $0.353$\\
    \#1 A & $373.15$ & $3300$ & $0.046$ & $0.352^*$ & $0.345$\\
    \#1 B & $373.15$ & $3300$ & $0.060$ & $0.345^*$ & $0.340$\\ 
    \#2   & $293.15$ & $3300$ & $0.250$ & $0.330$ & $0.319$\\
    \#3   & $293.15$ & $8000$ & $0.120$ & $0.320$ & $0.324$\\
	\bottomrule
	\end{tabularx}
	\label{tab_tones}
\end{table}

\subsection{Disqualification of screech}

Having a plate distance large enough, so that the relevant shock cells for the screech feedback loop (number three to five) fit into the domain, it is possible that screech noise is radiated by the impinging jet. However, in \cite{WilkeSesterhenn2016} it is shown, that for the investigated configuration of a plate distance of $h/D=5$, screech is not the relevant sound source. An impinging jet was compared to a free jet with equal parameters, except the presence of the plate. The observed tonal noise of the impinging tones is no screech, since the mode and the frequency differs between the two configurations. The pressure and axial velocity profiles were compared. It was found that the spacing of the first five shock cells does not differ between the two cases. This means, if the observed tone of the impinging jet were screech, it would need to have the exact same frequency like the tone of the free jet. Since this is not the case, screech can be excluded as the reason of tonal noise of the impinging jet. The sound pressure level of the impinging jet is more than 20 dB higher than the one of the free jet with the same parameters in the relevant frequency range $0.2 \leq Sr \leq 1$. Screech may exist additionally, but cannot be observed since the impinging tones are of much stronger nature and raise the ground sound pressure level above the screech peak of the free jet.

\subsection{Zone of silence}

As stated in the introduction, a hypothesis explaining the sound source mechanism according to Sinibaldi et al. \cite{SinibaldiMarino2015} can be summarised as follows: In the pre-silence region no standoff shock is present. Vortices interact directly with the impinging plate (direct shear layer-plate interaction). In the post-silence region the standoff shock disturbs the vortex-wall-interaction. The tones are only related to strong oscillations of the standoff shock. In the zone of silence, a smooth change between those two behaviours is observed.

The presently described simulations with $h/D=5$ and $NPR=2.15$ are located in the pre-silence zone. However we clearly observe standoff shocks in the numerical data. As described in the previous sections, the impinging tones are not caused by direct vortex-plate interactions but rather due to shock-vortex- or shock-vortex-shock-interactions. The observation of standoff shocks in the pre-silence zone is supported by the experiments of Buchmann et al.: in \cite{BuchmannMitchellSoria2011}, figure 2, schlieren images are shown for such a case ($h/D=4, NPR=3.2$) with present standoff shocks.

A hypothesis that explains the observations can be formulated as follows: Standoff shocks are present in both the pre- and the post-silence zone. However those shocks differ. In the pre-silence zone there is enough space for the jet shock cell system to damp before the flow reaches the impinging plate. Therefore the shocks can appear, disappear and move between the wall and the shock cell system. Those moving shocks are therefore difficult to detect in statistical values like root-mean-squares of velocity fluctuations. However, they can be observed using DNS or schlieren. In the post-silence zone, the impinging wall is directly located in the strong shock cells of the free jet and form a quasi-stationary system. Therefore they can be detected more easily in statistical data.

%
%
%
%
%
%
%

\section{Conclusion}
\label{sec:conclusion}

Despite the general accordance that impinging tones are produced due to a feedback loop, inconsistent statements about the production of the sound waves can be found in literature. In addition, no consensus could be found if standoff shocks are present in the pre-silence zone.

In order to clarify the open questions, we performed direct numerical simulations with a nozzle pressure ratio of 2.15 and a nozzle-to-plate distance of five diameters at Reynolds numbers of 3300 and 8000. Analysing the data, we find that standoff shocks periodically appear, disappear and move between the impinging plate and the shock cell system. Multiple standoff shocks can exist simultaneously, usually two or three are present for the chosen set of parameters. Concerning the generation of impinging tones, we clearly observe the feedback loop and prove that the interaction between vortices and standoff shocks produce the sound waves via two different mechanisms. One of the two mechanism can analogously be found in free jets and is responsible for screech. The difference however is that not the shock diamonds, but the standoff shock is involved in the interaction with the vortices. The impinging tone is not related to screech. The mode of the impinging jet is axisymmetrical.

\section{Acknowledgments}
The simulations were performed on the national supercomputer Cray XE6 (Hermit) and Cray XC40 (Hornet, Hazelhen) at the High Performance Computing Center Stuttgart (HLRS) under the grant number GCS-ARSI/44027.

The authors gratefully acknowledge support by the Deutsche Forschungsgemeinschaft (DFG) as part of the collaborative research center SFB 1029 "Substantial efficiency increase in gas turbines through direct use of coupled unsteady combustion and flow dynamics".


\section*{References}
\bibliography{0_literatur_impjet}

\begin{thebibliography}{10}
\bibitem{Neuwerth1974}
G.~Neuwerth, Acoustic feedback of a subsonic and supersonic free jet which
  impinges on an obstacle, NASA TTF-15719.
\end{thebibliography}

\end{document}